\documentstyle[12pt]{article}
 \makeatletter \textheight 24cm
\newcommand{\nn}{\nonumber}
\newcommand{\bd}{\begin{document}}
\newcommand{\ed}{\end{document}}
\newcommand{\bc}{\begin{center}}
\newcommand{\ec}{\end{center}}
\newcommand{\be}{\begin{eqnarray}}
\newcommand{\ee}{\end{eqnarray}}
\renewcommand{\thefootnote}{\alph{footnote}}
\newcommand{\se}{\section}
\newcommand{\sse}{\subsection}
\newcommand{\bi}{\bibitem}

\def\figcap{\section*{Figure Captions\markboth
     {FIGURECAPTIONS}{FIGURECAPTIONS}}\list
     {Figure \arabic{enumi}:\hfill}{\settowidth\labelwidth{Figure 999:}
     \leftmargin\labelwidth
     \advance\leftmargin\labelsep\usecounter{enumi}}}
\let\endfigcap\endlist \relax
\def\reflist{\section*{References\markboth
     {REFLIST}{REFLIST}}\list
     {[\arabic{enumi}]\hfill}{\settowidth\labelwidth{[999]}
     \leftmargin\labelwidth
     \advance\leftmargin\labelsep\usecounter{enumi}}}
\let\endreflist\endlist \relax

\begin{document}
\tolerance=10000 \baselineskip=7mm

\begin{titlepage}

 \vskip 0.5in
 \null
\begin{center}
 \vspace{.15in}
{\LARGE {\bf Study of Rare $B_{c}^{+} \to D_{d,s}^{(*) +}l
~\bar{l}$ Decays} }
\\ \vspace{1.0cm}
\par
 \vskip 2.1em
 {\large
  \begin{tabular}[t]{c}
{\bf C.Q. Geng$^{a,b}$, C.W. Hwang$^{a}$, and C.C. Liu$^{a,b}$}
\\
\\
{\sl $^a$Department of Physics, National Tsing Hua University}
\\  {\sl  $\ $ Hsinchu, Taiwan, Republic of China }
\\
\\
{\sl $^b$Theory Group, TRIUMF}
\\
{\sl  $\ $  4004 Wesbrook Mall, Vancouver, B.C. V6T
2A3, Canada}
\\

   \end{tabular}}
 \par \vskip 5.3em

\date{\today}
 {\Large\bf Abstract}
\end{center}

We study the rare decays of $B_{c}^{+} \to D_{q}^{(*)
+}l ~\bar{l}~$ ($q=d,s$ and $l=\nu_l,e,\mu,\tau)$ in the standard model.
The form factors are evaluated in the light front and constituent
quark models, respectively.
We find that the decay branching ratios calculated in the two models
for $B_{c}^{+}\to D^+_{q}l ~\bar{l}$
agree well with each other, whereas those for
$B_{c}^{+}\to D_{q}^{*+}l ~\bar{l}$ are different.

\end{titlepage}

\section{Introduction}

Recently, the CDF Collaboration has observed  the bottom-charm $B_c$
meson at Tevatron in Fermilab  \cite{cdf,PDG}. Its
 mass and lifetime are given as
$M_{B_c}=6.40\pm 0.39\ GeV$ and $\tau_{B_c} =(0.46^{+0.18}_{-0.16})\times
10^{-12}\ s$, respectively. The study of the $B_c$ meson is quite
interesting due to the following four main reasons:
(i)
$B_c$ is the lowest bound state of two heavy quarks ($b$ and $c$)
with open (explicit) flavor. It can be compared with the hidden
(implicit) flavor ($\bar{c} c$)
charmonium and ($\bar{b} b$) bottomonium.
 The hidden-flavor states decay strongly and electromagnetically
whereas the $B_c$ meson does weakly because it is below the
$B\bar{D}$-threshold.
(ii)
One may expect that the weak decays
of the $B_c$ meson are similar to those of  $B_{u,d,s}$ mesons.
However, the major difference between the weak decay properties of
$B_c$ and $B_{u,d,s}$ is that those of the latter ones are
described very well in the framework of the heavy quark limit. In
this limit the weak decay form factors are blind to the flavor and
spin orientation of the heavy quark. All of them can be expressed
through a single Isgur-Wise function \cite{IW}. In the
case of $B_c$, the heavy flavor and spin symmetries must be
reconsidered because both $b$ and $c$ quarks are heavy. Thus the
study with the finite quark mass is a more appropriate way.
(iii)
There have been many investigations of rare radiative, leptonic and
semileptonic decays of $B_{u,d,s}$ mesons induced by the flavor-changing
neutral current transitions of $b \to s,d$ \cite{Bdecays}
since the CLEO observation \cite{cleo} of
$b\to s\gamma $.
More recently, the process of $B\to K\mu^+\mu^-$ has been also
seen \cite{Belle} at the Belle detector in the KEKB $e^+e^-$ storage
ring.
In the standard
model (SM), these transitions are forbidden at the tree
level and occur only through loop diagrams.
 The studies are even more complete if similar decays for
$B_c$ are also included, which can be achieved by
introducing the spectator quark of $c$ in the diagrams.
In fact, some of the works
have been done and they can be
found in Refs. \cite{Bclist,Const,Du,Geng}.
(iv)
It is believed that there are about $10^8-10^9\ B_c$ mesons to be produced
in  future experiments at hadronic colliders \cite{Bcpro}, such as
the BTeV and LHC-B experiments \cite{Bexpts}. In these experiments,
most of rare $B_c$ decays should be accessible.

In this paper, we will concentrate on the rare decays of
$B_c^+\to D_q^{(*)}l\bar{l}\ (q=d,s)$ due to the $b \to q$ transitions
as shown in Figure 1 in the SM, which have not yet been explored in the
literature.
To study their decay rates and branching ratios,
we need to calculate the transition form factors
of the vector, axial-vector, and tensor currents, which
must be treated with
the non-perturbative method. There are many different
candidates for this purpose, {\em e.g.}, lattice QCD
\cite{9710057}, QCD sum rule \cite{sumrules1,sumrules2}, and
phenomenological models. In this work, we use the frameworks of two
phenomenological models: the light front quark model
(LFQM) \cite{hwcw,tensor} and the constituent quark model (CQM)
\cite{DM:D57,DM:D62}, to evaluate the form factors.

This paper is organized as follows. In Sec. 2,
we calculate
the form factors for $B_c^+\to D_{d,s}^{(*)+}$ transitions
in the LFQM and CQM.
 In Sec. 3, we study the differential
rates and branching ratios of $B_c^+\to P\l\bar{l}$ and $B_c^+\to
V l\bar{l}$
decays with $l=\nu,e,\mu,\tau$ and $P (V)=$ pseudoscalar
(vector) meson, respectively.
We also compare the results in the two models.
 Our conclusions are given in Sec. 5.
\section{Formalism and Models}

\subsection{Matrix Elements}

To get the transition matrix elements of $B_c^+\to P(V)$ with various
quark models, we parametrize them in terms of the relevant form
factors as follows:
\begin{eqnarray}
 \left\langle P(p_{2})|\ V_\mu \ |B_c(p_{1})\right\rangle
      &=& F_{+}(q^{2}) P_{\mu }+F_{-}(q^{2}) q_{\mu }\,,
      \nonumber \\
 \left\langle P(p_{2}) |\ T_{\mu\nu} q^\nu \ |B_c(p_{1}) \right\rangle
      &=& \frac{1}{m_{B_c}+m_{P}}\left[ q^{2}P_{\mu }-\left( P\cdot
q\right) q_{\mu }\right] F_{T}(q^{2}) \,,
\nonumber \\
 \left\langle V(p_{2},\epsilon ) |\ V_\mu \mp A_\mu \ |B_c(p_{1})
         \right\rangle &=&\frac{1}{m_{B_c}+m_{V}}
         \left[
-iV(q^{2}) \varepsilon _{\mu \nu \alpha \beta }
         \epsilon ^{*\nu }P^{\alpha }q^{\beta }
 \right.
 \nonumber \\
&&\pm A_{0}(q^{2})\left(P\cdot q\right) \epsilon _{\mu }^{*}
\pm  A_{+} (q^{2}) (\epsilon ^{*}\cdot P)P_{\mu }
 \nonumber \\
&& \left.
\pm A_{-}(q^{2}) (\epsilon ^{*}\cdot P)q_{\mu }
\right] \,,
\nonumber \\
\left\langle  V(p_{2},\epsilon ) |\ (T_{\mu\nu} \pm T^5_{\mu\nu})q^\nu \
|B_c(p_{1}) \right\rangle
         &=&-ig(q^{2}) \varepsilon _{\mu \nu \alpha \beta } \epsilon
         ^{*\nu }P^{\alpha }q^{\beta } \nonumber \\
      &&\pm a_{0}(q^{2}) \left( P\cdot q\right) \left[ \epsilon _{\mu
}^{*}
         -\frac{1}{q^{2}}\left( \epsilon^{*}\cdot q \right) q_{\mu }\right]
 \nonumber \\
      &&\pm a_{+}(q^{2})(\epsilon ^{*}\cdot P)
         \left[ P_{\mu }-\frac{1}{q^{2}}\left( P\cdot q\right)q_{\mu }\right]
 \,,
\label{defo}
\end{eqnarray}
where $m_i\ (i=B_c,P,V)$ are the meson masses, $p_1 (p_2)$ is the
momentum of the initial (final) meson, $\epsilon$ is the vector
meson polarization vector, $P=p_1+p_2$, $q=p_1-p_2$, $V_\mu=
\bar{q}_{2}\gamma _{\mu } q_{1}$, $A_\mu=\bar{q}_{2}\gamma _{\mu }
\gamma_5 q_{1}$, $T_{\mu\nu}=\bar{q}_{2}i\sigma _{\mu \nu }
q_{1}$, $T^5_{\mu\nu}=\bar{q}_{2}i\sigma _{\mu \nu }\gamma_5
q_{1}$, and $F_{\pm,T}$, $V$, $A_{0,\pm}$, $g$, and $a_{0,\pm}$
are the form factors.

Since the calculations of the transition form factors in Eq.
(\ref{defo}) belong to the nonperturbative regime, the
phenomenological quark models may be needed. One thing worthwhile
mentioning here is that all of form factors will be studied in the
time-like physical meson decay region of $0\leq q^2 \leq
(m_{B_c}-m_{P(V)})^2$. As  $q^2$ decreases (corresponding to the
increasing recoil momentum), we have to start considering
relativistic effects seriously. In particular, at the maximum
recoil point of $q^2=0$ where the final meson could be highly
relativistic, there is no reason to expect that the
non-relativistic quark model is still applicable. A consistent
treatment of the relativistic effects of the quark motion and spin
in a bound state is a main issue of the relativistic quark model.

\subsection{LFQM}

The LFQM \cite{Ter,Chung} is the relativistic quark model in which
a consistent and fully relativistic treatment of quark spins and
the center-of-mass motion can be carried out. This model has many
advantages. For example, the light-front wave function is
manifestly Lorentz invariant as it is expressed in terms of the
momentum fraction variables (in ``+" components) in analog to the
parton distributions in the infinite momentum frame. Moreover,
hadron spin can also be correctly constructed by using the so-called
Melosh rotation. The kinematic subgroup of the light-front
formalism has the maximum number of interaction-free generators
including the boost operator which describes the center-of-mass
motion of the bound state (for a review of the light-front
dynamics and light-front QCD, see Ref. \cite{Zhang}).

The LFQM has been applied to study the heavy-to-heavy
and heavy-to-light weak decay form factors in the timelike region
\cite{hwcw,time1}. These calculations are based on the observation
\cite{Dubin} that in the frame where the momentum transfer is
purely longitudinal, i.e., $q_\perp=0$, $q^2=q^+q^-$ covers the
entire range of momentum transfers. The price one has to pay is
that, besides the conventional valence-quark contribution, one
must also consider the non-valence configuration (or the so-called
$Z$ graph) arising from the quark-pair creation in the vacuum.
Unfortunately, a reliable way of estimating the $Z$ graph is still
lacking. However, the non-valence contribution vanishes if
$q^+=0$, and it is supposed to be unimportant for heavy-to-heavy
transitions \cite{hwcw}. In this paper, all of the values
obtained from the LFQM are based on the formulas in Refs.
\cite{hwcw,tensor}. We note that the form factors in Eq.
(\ref{defo}) depend on the meson $(H=q_1\bar{q}_2$) wave functions 
$\Phi_H$.
To fix the parameters in the wave functions, one may use
the meson decay constants $f_H$, given by
\be
f_H &=& \sqrt{24}
\int {dx\,d^2k_\perp\over 
2(2\pi)^3}\,\Phi_H(x,
k_\perp)\,{{\cal A}\over\sqrt{{\cal A}^2+k_\perp^2}}, 
\label{fp}
\ee
where ${\cal A}=m_{q_1}x+m_{q_2}(1-x)$ with $m_{q_i}$ being the quark 
masses 
and $\vec{k}_\perp$ is the component of the internal momentum 
$\vec{k}=(\vec{k}_\perp,k_z)$.

\subsection{CQM}

As mentioned in Sec. 1, there are also other theoretical
approaches for calculating the form factors. However,
the theoretical uncertainties are large and each of these methods
has only a limited range of applicability. For example,
the model with QCD sum rules gives good results
for the form factors at the low $q^2$ region; whereas
the lattice QCD is appropriate only at
the high $q^2$ one. In spite of that the quark models
can be used to evaluate the form factors in the full $q^2$ range,
they are not closely related to the QCD Lagrangian and have many
input parameters which are not measurable directly. Therefore, a
relativistic constituent quark model is suggested in Ref.
\cite{DM:D53} which combines several theoretical methods such as
the constituent quark models, QCD sum rules,
lattice QCD calculations, and analytical constrains. This model used the
light-cone technique with the relativistic double spectral representations
in the initial and final meson wave functions.
Explicitly, they calculated
the form factors at $q^2<0$, i.e. the space-like
region, by choosing $P_\bot =0$, $q_+=0$, and $q_\bot ^2=-q^2$.
In order to obtain the form factors in the $q^2>0$ region,
in Ref. \cite{DM:D53}, some modifications from
the space-like formulas were used to get their values in
$0<q^{2}<(m_{b}-m_{d,s})^{2}$.
It is known that in the time-like region
$q^2>0$, there are the normal and anomalous parts,
 respectively.
The result for the former is the same as  that for $q^2<0$,
but for the latter it can be ignored for small $q^2>0$
and rises sharply as $q^2\to (m_b-m_{d,s})^2$.

In this paper, we will evaluate the form factors of $B_c^+\to
D^{(*)+}_{d,s}$ in the CQM by using the results in Refs.
\cite{DM:D57,DM:D62}. In the calculations, 
we first compute the values for the normal
parts in $0<q^2<(m_{b}-m_{d,s})^{2}$ and then fit the data 
in terms of the double pole form, given by
\be
F_i(q^2)=\frac{F_i(0)}{~~1+\sigma_1s+\sigma_2s^2~~} 
\label{Fit}
\ee where
$s=q^2/m_{B_c}^2$, $F_i(0)$ are the form factors at $q^2=0$, and
$\sigma _{1,2}$ are the fitted parameters. 
The form factors in the remaining regions of
$(m_{b}-m_{d,s})^2\leq  q^2 \leq (m_{B_c}-m_{P,V})^2$
can be extrapolated from Eq. (\ref{Fit}).

\subsection{Form factors}

As in Refs. \cite{hwcw,tensor,DM:D57,DM:D62},
in this paper we choose the Gaussian-type meson wave function
 for both LFQM and CQM to calculate the form factors, i.e.,
\begin{equation}
\Phi_H\propto \exp(-\frac{\vec{k}^2}{2\omega_H ^2})\,,
 \label{Gaus}
\end{equation}
where $\vec{k}$ and $\omega_H$ are the internal momentum and the  scale
parameter of $H$ meson, respectively.

To find the numerical values of the form factors in the two models,
we need to specify the parameters appearing in the wave functions.
In the LFQM, we use the decay constants to constrain the quark
masses and $\omega_H$ in Eq. (\ref{Gaus}) \cite{hwcw}. However,
since the decay constants of heavy mesons are unknown
experimentally,
 we have to rely on results in other QCD models such as  the lattice
QCD. Explicitly, we take \cite{PDG,Const}
\begin{eqnarray}
&&f_{B_c}=360~{\rm MeV},~~~~~f_{D_d}=200~{\rm MeV},~~~~~
f_{D^*_d}=250~{\rm MeV},\nonumber \\ &&f_{D_s}=230~{\rm
MeV},~~~~~f_{D^*_s}=330~{\rm MeV},~~~~~m_d=0.25~{\rm
GeV},\nonumber \\ &&m_s=0.40~{\rm GeV},~~~~~m_c=1.60~{\rm
GeV},~~~~~m_b=4.80~{\rm GeV},
 \label{241}
\end{eqnarray}
which fix the scale parameters to be
\begin{eqnarray}
&&\omega_{B_c}=0.81~{\rm GeV},~~~~~\omega_{D_d}=0.46~{\rm
GeV},~~~~~\omega_{D^*_d}=0.47~{\rm GeV},\nonumber \\
&&\omega_{D_s}=0.50~{\rm GeV},~~~~~\omega_{D^*_s}=0.56~{\rm
GeV}\,,
\label{242}
\end{eqnarray}
respectively. 
In our calculations, we also take $m_{B_c}= 6.4\
GeV$ and $\tau_{B_c}= 0.46\times 10^{-12}\  s$. 
In order to compare the numerical values in the LFQM and CQM, 
we shall use the same decay constants, quark masses and scale parameters 
in both models.
We note that in the LFQM, the light quark masses in Eq. (\ref{241})
are fixed by using the kaon decay constant $f_K=159.8\ MeV$ and charge 
radius $<r_K^2>=0.34\ fm^2$, while in both models a different set of the 
heavy quark masses has a little effect on the form factors.

 Based on the parameters in Eqs. (\ref{241}) and (\ref{242}),
we show the $q^2$ dependences of the
form factors for $B_c^+\to D^{(*)+}$ and $B_c^+\to D^{(*)+}_s$
in Figures 2 and 3,
respectively.
The numerical results for the form factors at
$q^2=0$ are listed in Table 1.
From the table, we see that the values of the form factors
at $q^2=0$ in the LFQM and CQM agree well with each other
except $A_{\pm}(0)$.
However, as shown in Figures 2 and 3, the results at large $q^2$
in the two models are quite different.

\begin{table}[h]
   \caption{Form factors for $B_{c}^{+}\rightarrow D_{d,s}^{(*)+}$
    transitions at $q^{2}=0$ in LFQM and CQM models, where $F_{\pm,T}$
are for $B_c^+\to P^+\ (P=D,D_s)$ and
$V,A_{0,\pm},g$ and $a_{0,+}$ for $B_c^+\to V^+\ (V=D^*,D^*_s)$,
respectively.}
   \label{form1}
   \begin{center}
   \begin{tabular}{|r|r|r|r|r|r|r|r|r|}
   \hline
   & \multicolumn{4}{c|} {$B_{c}^{+}\rightarrow D^{(*)+}$} & \multicolumn{4}{c|}
   {$B_{c}^{+}\rightarrow D_{s}^{(*)+}$} \\ \hline
   & LFQM & \multicolumn{3}{c|}{CQM}& LFQM & \multicolumn{3}{c|}{CQM}
   \\ \hline
   & $F_i(0)~$ & $F_i(0)~$&$\sigma_1~~$&$\sigma_2~$& $F_i(0)~$& $F_i(0)~$&$\sigma_1~~$&$\sigma_2~$\\ \hline
   $F_{+}$ & $0.126$ & $0.123$ &$-3.35$&$3.03$& $0.165$ & $~0.167$ & $-3.40$ & $3.21$\\ \hline
   $F_{-}$ & $-0.141$ & $-0.130$ &$-3.63$&$3.55$& $-0.186$ & $-0.166$ & $-3.51$ & $3.38$\\ \hline
   $F_{T}$ & $-0.199$ & $-0.186$ &$-3.52$&$3.38$& $-0.258$ & $-0.247$ & $-3.41$ & $3.30$\\ \hline
   $V$     & $-0.208$ & $-0.198$ &$-3.63$&$3.65$& $-0.336$ & $-0.262$ & $-3.49$ & $3.51$\\ \hline
   $A_{0}$ & $-0.198$ & $-0.198$ &$-2.81$&$2.53$& $-0.330$ & $-0.280$ & $-2.66$ & $2.24$\\ \hline
   $A_{+}$ & $0.079$ & $0.108$ &$-3.12$&$2.94$& $0.118$ & $0.144$ & $-2.99$ & $2.95$\\ \hline
   $A_{-}$ & $-0.098$ & $-0.185$ &$-3.45$&$3.54$& $-0.130$ & $-0.246$ & $-3.34$ & $3.46$\\ \hline
   $g$     & $0.130$ & $0.124$ &$-3.63$&$3.65$& $0.214$ & $0.167$ & $-3.45$ & $3.29$\\ \hline
   $a_{0}$ & $0.130$ & $0.124$ &$-2.82$&$2.53$& $0.214$ & $0.167$ & $-2.63$ & $2.23$\\ \hline
   $a_{+}$ & $-0.130$ & $-0.124$ &$-3.31$&$3.14$& $-0.214$ & $-0.167$ & $-3.16$ & $3.13$\\ \hline
   \end{tabular}
   \end{center}
\end{table}

\section{Decay rates and branching ratios}

In the SM, the contributions to the rare decays of $B_c^+\to
D_{d,s}^{(*)+}l\bar{l}$ arise from the $W$-box
and $Z(\gamma )$-penguin diagrams as seen in Figure 1.
The effective Hamiltonians of $b\rightarrow  q\nu  \bar{\nu }$
$(q=s,d)$ are given by \cite{IL}

\begin{equation}
    {\cal H}=\frac{G_{F}}{\sqrt{2}} \frac{\alpha _{em}}{2\pi sin^{2}\theta _{W}}
     \lambda _{t} D\left( x_{t}\right) \bar{b}\gamma _{\mu }\left( 1-\gamma
     _{5}\right) q\bar{\nu }\gamma _{\mu}\left( 1-\gamma _{5}\right) \nu
\label{Hnn}
\end{equation}
where $G_{F}$ is the Fermi constant, $x_{t}\equiv
m_{t}^{2}/m_{W}^{2}$, $\lambda _{t}=V_{tb}V_{tq}^{*}$ is the
product of the CKM elements, and the $m_t$ dependent function of
$D\left( x_{t}\right)$ can be found in Refs. \cite{BG,BU}.

The effective Hamiltonians of $b\rightarrow q l^{+}l^{-}\ (q=s,d)$
are given by \cite{IL}
\be
    {\cal H}=\frac{G_{F}\alpha _{em}\lambda _{t}}
{\sqrt{2}\pi }\left[ C_{8}(\mu)
    \bar{s}_{L}\gamma _{\mu }b_{L}\ \bar{l}\gamma ^{\mu }l
    +C_{9} \bar{s} _{L}\gamma _{\mu }b_{L}\ \bar{l}\gamma ^{\mu }\gamma _{5}l
    -\frac{ 2m_{b}C_{7}(\mu) }{q^{2}}\bar{s}_{L}i\sigma _{\mu \nu }
    q^{\nu }b_{R}\ \bar{l}\gamma ^{\mu }l\right]
\label{Hll}
\ee
where $C_{8}(\mu )$, $C_{9}$ and $C_{7}(\mu )$ are Wilson coefficients 
(WCs) and 
their expressions can be found in Ref. \cite{BU} for the SM. We note that
$C_{9}$ is free of the 
$\mu $ scale. Besides the short-distance (SD)
contributions, the main effect on the decays is from c\={c}
resonant states such as $\Psi$ and $\Psi ^{\prime }$, $i.e.$, 
the long-distance (LD) contributions. To including the LD effect, in Eq. 
(\ref{Hll}) we
 replace $C_{8}(\mu)$ by $C_{8}^{eff}(\mu)$ \cite{BU,Chen}, given by
\begin{equation}
C_{8}^{eff}(\mu)=C_{8}( \mu) +\left( 3C_{1}\left( \mu \right)
+C_{2}\left( \mu \right) \right) \left( h\left( x,s\right) 
+\frac{3}{\alpha }%
\sum_{j=\Psi ,\Psi ^{\prime }}k_{j}\frac{\pi \Gamma \left( j\rightarrow
l^{+}l^{-}\right) M_{j}}{q^{2}-M_{j}^{2}+iM_{j}\Gamma _{j}}\right) \,, 
\label{effc8}
\end{equation}
where we have neglected the small WCs, and $h(x,s)$ describes
the one-loop matrix elements of operators $O_{1}=\bar{s}_{\alpha }\gamma
^{\mu }P_{L}b_{\beta }\ \bar{c}_{\beta }\gamma _{\mu }P_{L}c_{\alpha }$ 
and $%
O_{2}=\bar{s}\gamma ^{\mu }P_{L}b\ \bar{c}\gamma _{\mu }P_{L}c$ 
\cite{BU}, 
$M_{j}$ ($\Gamma _{j}$) are the masses (widths) of intermediate states,
and  $k_{j}=-1/\left(
3C_{1}\left( \mu \right) +C_{2}\left( \mu \right) \right) $ \cite{Chen}.

 From Eqs. (\ref{Hnn}) and
 (\ref{Hll}),
the differential decay rates
for $B_c^+\to H l \bar{l }\ (H=P
 ,V)$
are found to be \cite{Wyler,Kao}

\be
    \frac{d\Gamma \left( B_c^+\rightarrow P \nu \bar{\nu }\right) }{ds}
       &=&\frac{G_F^2m_{B_c}^5|\lambda _t|^2
       \alpha _{em}^2|D\left( x_{t}\right) |^2}{2^8\pi ^5
       sin^4\theta_W}\left|F_+\right|^2\phi_H^{3\over 2} \,, \label{Rate1}
\\
\nn\\
    \frac{d\Gamma \left( B_c^+\rightarrow V \nu \bar{\nu }\right) }{ds}
       &=&\frac{3G_F^2m_{B_c}^5|\lambda _t|^2\alpha _{em}^2|D\left( x_t\right) |^2}
       {2^8\pi ^5sin^4\theta_W}
       \phi_H^{1\over 2}\left[ s\alpha _1+ {\phi_H\over 3}
       \beta _1\right]
       \,, \label{Rate2}
\\
\nn\\
    \frac{d\Gamma \left( B_c^+\rightarrow P l^+l^-\right) }{ds}
        &=&\frac{G_F^2|\lambda _t|^2m_{B_c}^5\alpha _{em}^2}
        {3\cdot 2^9\pi ^5}v\phi_H ^{1\over 2}\left[ \left(1+\frac{2t}
        {s}\right) \phi_H \alpha _2+12t \beta _2\right]\,,
\label{Rate3}
\ee
and
\be
    \frac{d\Gamma \left( B_c^+\rightarrow Vl^{+}l^{-}\right) }{ds}
       &=&\frac{G_F^2m_{B_c}^5|\lambda _t|^2\alpha _{em}^2}{2^9\pi ^5}
       v\phi_H ^{1\over 2} \left[
       \left( 1+\frac{2t}{s}\right) \left(
       s\alpha _3+\frac{\phi _H}{3}\beta _3 \right)
       +4t\delta \right],
\label{Rate4}
\ee
respectively,
where $s = q^2/m_{B_c}^2$,
    $t= m_l^2/ m_{B_c}^2$,
    $r_H=m_H^2/m_{B_c}^2$,
    $v=\sqrt{1-4t/ s}$,
 and
the expressions of
$\phi_H$, $\alpha_i$, $\beta_i$ \cite{Wyler} and $\delta$ \cite{Kao}
are given in Appendix.


By using the form factors of the LFQM and CQM in Figures 2 and 3,
Eqs. (\ref{Rate1})-(\ref{Rate4}), and
$|\lambda_t|=|V_{tb}V_{tq}|=0.041\,
(0.008)$ for $q=s\,(d)$ \cite{Buras01},
we now estimate the numerical
values of the decay
rates for $B_{c}^{+} \to D_{d,s}^{(*) +}\nu \bar{\nu}$ and
$B_{c}^{+} \rightarrow D_{d,s}^{(*)+}l^+l^-$.
Our results for the
differential decay branching ratios as a function of $s$ are shown
in Figures 4-9, respectively. Here, for the charged lepton modes,
we have presented our studies both with and without long-distance
contributions. We note that the results for the electron modes are
the same as the corresponding muon ones. We also note that the at
the large $q^2$ region, all the rates in the figures decrease
because  $\phi _H$ go to zero as $q^2\to
(m_b-m_{d,s})^2$. We emphasize that all our numerical
predictions should be viewed as central values and
their errors depend on the uncertainties from the corresponding meson
decay constants and constituent quark masses as well as the CKM
parameters.

The decay branching ratios of $B_{c}^{+}
\rightarrow D_{d,s}^{(*)+}\nu \bar{\nu}$ and $B_{c}^{+}
\rightarrow
   D_{d,s}^{(*)+}l^{+}l^-\ (l=\mu,\tau)$
are summarized in Tables 2 and 3, respectively,
where LD effects for the charged lepton modes are not included.
With the LD effects, we
introduce some cuts close to $q^2=0$ and around the resonances
 of $J/\psi$ and $\psi^{\prime }$ and study
the three regions as follows
\be
   I: & & \ \ \ \ \ \ \ \ \sqrt{q^2_{min}}\;<\; \sqrt{q^2}\;<\; M_{J/\psi
}-0.20\,;
\nn\\
   II: & & M_{J/\psi}+0.04\;<\;\sqrt{q^2}\;<\;M_{\psi^{\prime}}-0.10\,;
\nn \\
   III: & & \ \ M_{\psi^{\prime}}+0.02\;
<\;\sqrt{q^2}\;<\;m_{B_c}-m_{P,V}\,, \label{Cuts} \ee where
$\sqrt{q^2_{min}}=2m_l$ and $0.5\ GeV$ for $B_c^+\to
D_{d,s}^+l^+l^-$ and $B_c^+\to D_{d,s}^{*+}l^+l^-$, respectively.
In Table 4, we present the decay branching ratios in terms of the
regions shown in Eq. (\ref{Cuts}).

\begin{table}[h]
   \caption{Decay branching ratios of
$B_{c}^{+} \rightarrow D_{d,s}^{(*)+}\nu\bar{\nu}$.}
   \label{br1}
   \begin{center}
   \begin{tabular}{|c|cc|}
   \hline
   ~ & LFQM & CQM \\ \hline
   $10^{8}$Br$( B_c^+\rightarrow D^+\nu \bar{\nu} ) $ & $ 2.77 $ & $ 2.74 $ \\
   $10^{8}$Br$( B_c^+\rightarrow D^{*+}\nu\bar{\nu}) $ & $ 7.64 $ & $ 5.99 $
\\ \hline
   $10^{6}$Br$( B_c^+\rightarrow D_s^+\nu\bar{\nu}) $ & $ 0.92 $ & $ 0.92 $ \\
   $10^{6}$Br$( B_c^+\rightarrow D_s^{*+}\nu\bar{\nu}) $ & $ 3.12 $ & $ 2.12 $
\\ \hline
   \end{tabular}
   \end{center}
\end{table}

\begin{table}[h]
   \caption{Decay branching ratios of
$B_{c}^{+} \rightarrow
   D_{d,s}^{(*)+}l^{+}l^{-}$ without including LD effects.}
   \label{br2}
   \begin{center}
   \begin{tabular}{|c|rr|}
   \hline
   & \multicolumn{2}{c|} {without LD}\\ \hline
   Decay Mode & LFQM & CQM \\ \hline
   $10^{8}$Br$( B_c^{+}\rightarrow D^{+}\mu ^{+}\mu ^{-}) $
                                & $ 0.41 $ & $ 0.40 $\\
   $10^{8}$Br$( B_c^{+}\rightarrow D^{*+}\mu ^{+}\mu ^{-}) $
                                & $ 1.01 $ & $ 0.79 $\\
   $10^{8}$Br$( B_c^{+}\rightarrow D^{+}\tau ^{+}\tau ^{-}) $
                                & $ 0.13 $ & $ 0.12 $\\
   $10^{8}$Br$( B_c^{+}\rightarrow D^{*+}\tau ^{+}\tau ^{-}) $
                                & $ 0.18 $ & $ 0.14 $\\ \hline
   $10^{7}$Br$( B_c^{+}\rightarrow D_{s}^{+}\mu ^{+}\mu ^{-})$
                                & $ 1.36 $ & $ 1.33 $\\
   $10^{7}$Br$( B_c^{+}\rightarrow D_{s}^{*+}\mu ^{+}\mu ^{-})$
                                & $ 4.09 $ & $ 2.81 $\\
   $10^{7}$Br$( B_c^{+}\rightarrow D_{s}^{+}\tau ^{+}\tau ^{-})$
                                & $ 0.34 $ & $ 0.37 $\\
   $10^{7}$Br$( B_c^{+}\rightarrow D_{s}^{*+}\tau ^{+}\tau ^{-})$
                                & $ 0.51 $ & $ 0.41 $\\ \hline
   \end{tabular}
   \end{center}
\end{table}

\begin{table}[h]
   \caption{Decay branching ratios of
$B_{c}^{+} \rightarrow
   D_{d,s}^{(*)+}l^{+}l^{-}$ with LD effects and the cuts.}
   \label{br3}
   \begin{center}
   \begin{tabular}{|c|rr|rr|rr|rr|}
   \hline
   & \multicolumn{8}{c|} {with LD}\\ \hline
   regions & \multicolumn{2}{c|} {I}
           & \multicolumn{2}{c|} {II}
           & \multicolumn{2}{c|} {III}
           & \multicolumn{2}{c|} {I+II+III} \\ \hline
   Decay Mode & LFQM & CQM & LFQM & CQM & LFQM & CQM & LFQM & CQM \\
\hline
   $10^{9}$Br$( B_c^{+}\rightarrow D^{+}\mu ^{+}\mu ^{-}) $
   & $1.48$ & $1.40$ & $0.75$ & $0.73$ & $1.09$ & $1.07$ & $3.31$& $3.20$ \\
   $10^{9}$Br$( B_c^{+}\rightarrow D^{*+}\mu ^{+}\mu ^{-}) $
   & $2.17$ & $1.55$ & $1.81$ & $1.49$ & $3.78$ & $2.95$ & $7.75$& $5.98$ \\
   $10^{9}$Br$( B_c^{+}\rightarrow D^{+}\tau ^{+}\tau ^{-}) $
   & $--$ & $--$ & $0.02$ & $0.01$ & $1.03$ & $0.94$ & $1.05$ & $0.95$ \\
   $10^{9}$Br$( B_c^{+}\rightarrow D^{*+}\tau ^{+}\tau ^{-}) $
   & $--$ & $--$ & $0.02$ & $0.02$ & $1.30$ & $1.02$ & $1.33$ & $1.03$ \\ \hline
   $10^{8}$Br$( B_c^{+}\rightarrow D_{s}^{+}\mu ^{+}\mu ^{-})$
   & $5.89$ & $5.83$ & $2.57$ & $2.47$ & $2.69$ & $2.66$ & $11.15$ & $10.96$ \\
   $10^{8}$Br$( B_c^{+}\rightarrow D_{s}^{*+}\mu ^{+}\mu ^{-})$
   & $11.90$ & $6.80$ & $8.30$ & $5.78$ & $11.18$ & $8.58$ & $31.38$ & $21.16$ \\
   $10^{8}$Br$( B_c^{+}\rightarrow D_{s}^{+}\tau ^{+}\tau ^{-})$
   & $--$ & $--$ & $0.05$ & $0.05$ & $2.67$ & $2.95$ & $2.72$ & $3.00$ \\
   $10^{8}$Br$( B_c^{+}\rightarrow D_{s}^{*+}\tau ^{+}\tau ^{-})$
   & $--$ & $--$ & $0.10$ & $0.08$ & $3.31$ & $2.73$ & $3.41$ & $2.80$ \\ \hline
   \end{tabular}
   \end{center}
\end{table}


As seen from Figures 4-9 and Tables 2-4,
the branching ratios of $B_c^+\to D^+_{d,s}l\bar{l}$  in
the LFQM and CQM are agree very well, while the results of
 $B_c^+\to D_{d,s}^{*+}l\bar{l}$ in the LFQM are larger
than those in the CQM but the
differences are at the $20\%$ level.

Finally, we remark that in our calculations on $B_c^+\to
D^{*+}_{q}l^+l^-\ (q=d,s)$, we have not included the contributions from
the weak annihilation accompanied by a photon emission which
are dominant in the decays of $B_c^+\to
D^{*+}_q\gamma$ \cite{Du}. However, they are only important at low $s$
and the cut at $\sqrt{q^2_{min}}$ in Eq. (\ref{Cuts}) should reduce
the contributions from the virtual photon diagrams.

\section{Conclusions}

We have studied the rare $B_c$ decays of
$B_{c}^{+} \to D_{d,s}^{(*) +}\nu \bar{\nu}$ and $B_{c}^{+} \to
D_{d,s}^{(*) +}l^+l^-$ ($l=e,\mu,\tau$).
In our analysis, we have used the form factors of
 $B_{c}^{+} \to D_{d,s}^{(*) +}$ transitions calculated in
the LFQM and CQM. We have found that $Br(B_{c}^{+} \to
D^+l\,\bar{l})\ (l=\nu,e,\mu,\tau)$= $(2.77, 0.41, 0.41, 0.13)$
and $(2.74, 0.40, 0.40, 0.12)\times 10^{-8}$, $Br(B_c^+\to D_s^+
l\,\bar{l})$=$(9.2, 1.36, 1.36, 0.34)$ and $(9.2, 1.33, 1.33, 0.37
)\times 10^{-7}$, $Br(B_{c}^{+} \to D^{*+} l\,\bar{l})$= $(7.64,
1.01, 1.01, 0.18)$ and $(5.99, 0.79, 0.79, 0.14)\times 10^{-8}$,
and $Br(B_{c}^{+} \to D^{*+}_s l\,\bar{l})$= $(31.2, 4.09, 4.09,
0.51)$ and $(21.2, 2.81, 2.81, 0.41)\times 10^{-7}$, in the two
models, respectively. Clearly, some of the above rare $B_c$ decays
can be measured at the BTeV and LHC-B experiments.

\section*{Acknowledgments}

This work was supported in part by the National Science Council of
the Republic of China under contract numbers
 NSC-90-2112-M-007-040.

\newpage

\bc
{\Large Appendix}
\ec

The parameters of
$\phi_H$, $\alpha_i$, $\beta_i\ (i=1,2,3)$ and $\delta $
in Eqs. (\ref{Rate1})-(\ref{Rate4})
are defined by

\be
    \phi_H &=&\left( 1-r_H\right) ^{2}-2s\left( 1+r_H\right) +s^{2},
     \\
\nonumber \\
    \alpha _1 &=& (1-\sqrt{r_H}) ^{2} \left| A_{0}\right| ^{2} +
           \frac{\phi_H
           }{(1+\sqrt{r_H})^{2}}\left| V\right| ^{2}
           \ ,
\nonumber \\
    \beta _1 &=& \frac{(1-\sqrt{r_H})^2}{4r_H} \left| A_{0}
           \right|^2 -\frac{s}{(1+\sqrt{r_H})^2}\left| V\right| ^2
           +\frac{\phi _H \left| A_+\right| ^2}{4r_H(1+\sqrt{r_H})^2}
           \nonumber \\
           &&+\frac{1}{2}\left( \frac{1-s}{r_H}-1\right)
           \frac{1-\sqrt{r_H}}{1+\sqrt{r_H}}Re(A_{0}A_{+}^*)\,,
\\
\nonumber \\
    \alpha _2 &=&
\left|C_{8}^{eff}F_{+}-\frac{2\hat{m}_bC_{7}F_{T}}{1+\sqrt{r_H}}\right|^{2}
       +|C_{9}F_{+}|^{2} ~,
 \nonumber \\
    \beta _2 &=& |C_{9}|^{2}\left[ \left(
       1+r_H-{s\over 2}\right) |F_{+}|^{2}+\left( 1-r_H\right)
       Re(F_{+}F_{-}^*)+\frac{1}{2}s|F_{-}|^{2}\right]\,,
\\
     \nonumber \\
  \alpha _3 &=& (1-\sqrt{r_H})^2\left[ \left| C_8^{eff}A_0
     -\frac{2\hat{m}_bC_7(1+\sqrt{r_H})a_0}{s}\right|^2
     +\left| C_9A_0\right| ^2\right]
     \nonumber \\
     &&+\frac{\phi _H}{(1+\sqrt{r_H})^2}\left[ \left| C_8^{eff}V
     -\frac{2\hat{m}_bC_7(1+\sqrt{r_H})g}{s}\right| ^2
     +\left| C_9V\right| ^2\right] \,,
     \nonumber \\
  \beta _3 &=& \frac{(1-\sqrt{r_H})^2}{4r_H}\left[ \left| C_8^{eff}A_0
     -\frac{2\hat{m}_bC_7(1+\sqrt{r_H})a_0}{s}\right| ^2
     +\left| C_9A_0\right| ^2\right]
     \nonumber \\
     &&-\frac{s}{(1+\sqrt{r_H})^2}\left[ \left| C_8^{eff}V
     -\frac{2\hat{m}_bC_7(1+\sqrt{r_H})g}{s}\right| ^2
     +\left| C_9V\right| ^2\right]
     \nonumber \\
     &&+\frac{\phi _H}{4r_H(1+\sqrt{r_H})^2}\left[ \left|
     C_8^{eff}A_+-\frac{2\hat{m}_bC_7(1+\sqrt{r_H})a_+}{s}\right| ^2
     +\left| C_9A_+\right| ^2\right]
     \nonumber \\
     &&+\frac{1}{2}\left(\frac{1-s}{r_H}-1\right)\frac{1-\sqrt{r_H}}{1+\sqrt{r_H}}
     Re\left\{ \left[
     C_8^{eff}A_0-\frac{2\hat{m}_bC_7(1+\sqrt{r_H})a_0}{s}\right]
     \right.
     \nonumber \\
     &&\left. \times \left[
     C_8^{eff}A_+-\frac{2\hat{m}_bC_7(1+\sqrt{r_H})a_+}{s}\right]
     +|C_9|^2Re(A_0A_+^*)\right\}
     \,,
\ee
and
\be
  \delta &=& \frac{|C_9|^2}{2(1+\sqrt{r_H})^2}\left\{ -2\phi
     _H|V|^2-3(1-r_H)^2|A_0|^2+\frac{\phi _H}{4r_H}\left[
     2(1+r_H)-s\right]|A_+|^2 \right.
     \nonumber \\
     &&\left. +\frac{\phi _Hs}{4r_H}|A_-|^2+\frac{\phi _H(1-r_H)}{2r_H}
     Re\left( A_0A_+^*+A_0A_-^*+A_+A_-^*\right) \right\}
     \,,
\ee
respectively, where $\hat{m}_b=m_b/m_{B_c}$.


\newpage
\begin{figcap}


\item
One-loop diagrams for the
short-distance contributions to the decays of $B_c^+\to
D_q^{(*)+}l\bar{l}\ (q=d,s)$ in the SM.

\item
 Form factors of (a)
$F_{\pm,T}$ for $B_c^+\to D^+$, and (b) $V$ and $A_0$, (c) $A_{\pm}$,
and (d) $g$ and $a_{0,+}$ for $B_c^+\to D^{*+}$. The solid and dashed
courves stand for the results from the LFQM and CQM,
respectively.

\item
Same as Figure 2 but replacing $D^{(*)}$ by $D_s^{(*)}$.

\item
 Differential decay
branching ratios as a function of $s=q^{2}/m_{B_c}^{2}$ for (a)
$B_c^+\rightarrow D^+ \nu\bar{\nu} $ and (b) $B_c\rightarrow D^{*}
\nu\bar{\nu} $. Legend is the same as Figure 2.

\item
 Same as Figure 4 but
for (a) $B_c^+\rightarrow D_s^+\nu\bar{\nu}$ and (b) $B_c^+\rightarrow
D_s^{+*} \nu\bar{\nu}$.

\item
 Same as Figure 4 but
for (a) $B_c^+\rightarrow D^+\mu ^{+}\mu ^{-}$ and (b)
$B_c^+\rightarrow D^+ \tau ^{+}\tau ^{-}$. The curves with and
without resonant shapes represent including and non-including LD
contributions, respectively.

\item
 Same as Figure 6 but
for (a) $B_c^+\rightarrow D_s^+ \mu ^{+}\mu ^{-}$ and (b)
$B_c^+\rightarrow D_s^+\tau^{+}\tau ^{-}$.

\item
 Same as Figure 6 but
for (a) $B_c^+\rightarrow D^{*+} \mu ^{+}\mu ^{-}$ and (b)
$B_c^+\rightarrow D^{*+} \tau ^{+}\tau ^{-}$.

\item
 Same as Figure 6 but
for (a) $B_c^+\rightarrow D_s^{*+} \mu ^{+}\mu ^{-}$ and (b)
$B_c^+\rightarrow D_s^{*+} \tau ^{+}\tau ^{-}$.

\end{figcap}

\newpage
\begin{figure}[h]
\includegraphics{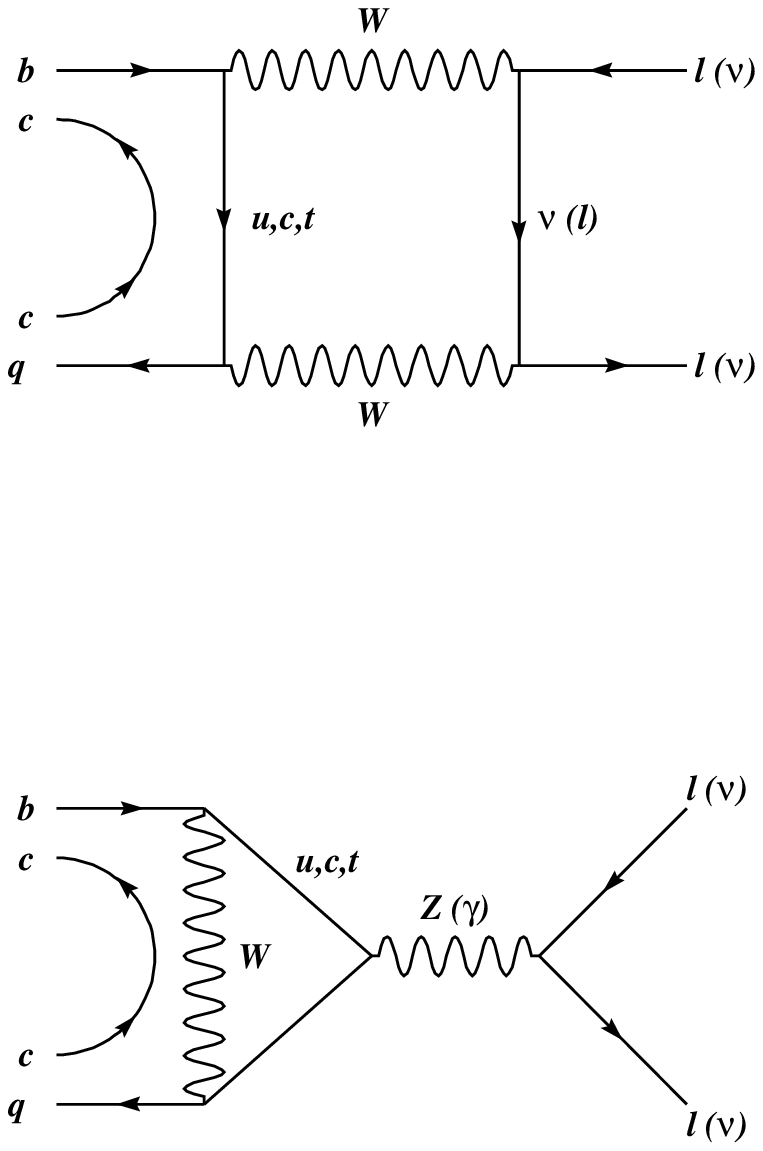} \vskip 13.cm \caption{One-loop diagrams for the
short-distance contributions to the decays of $B_c^+\to
D_q^{(*)+}l\bar{l}\ (q=d,s)$ in the SM.} \label{Feynman}
\end{figure}

\newpage
\begin{figure}[h]
\includegraphics{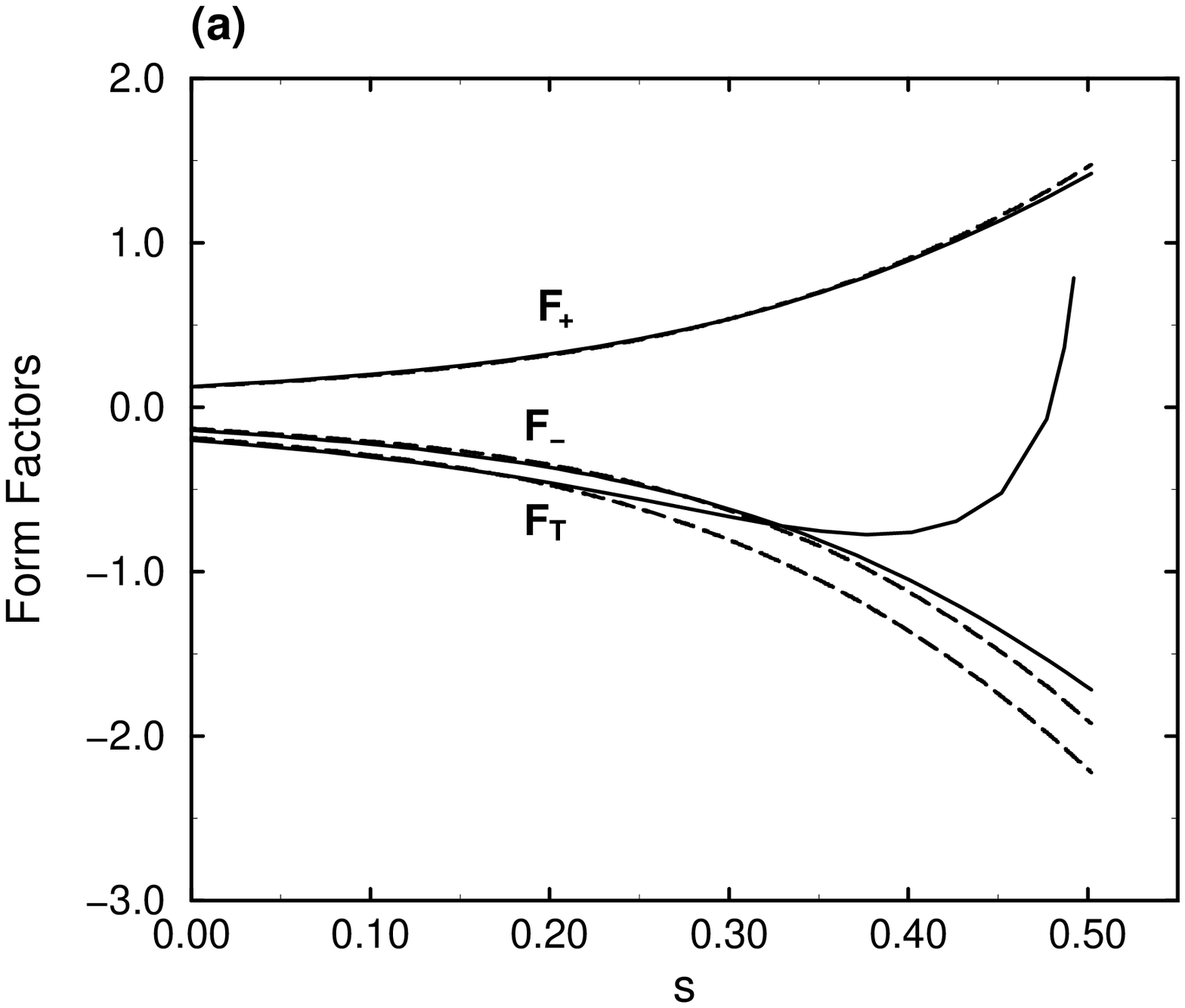} \includegraphics{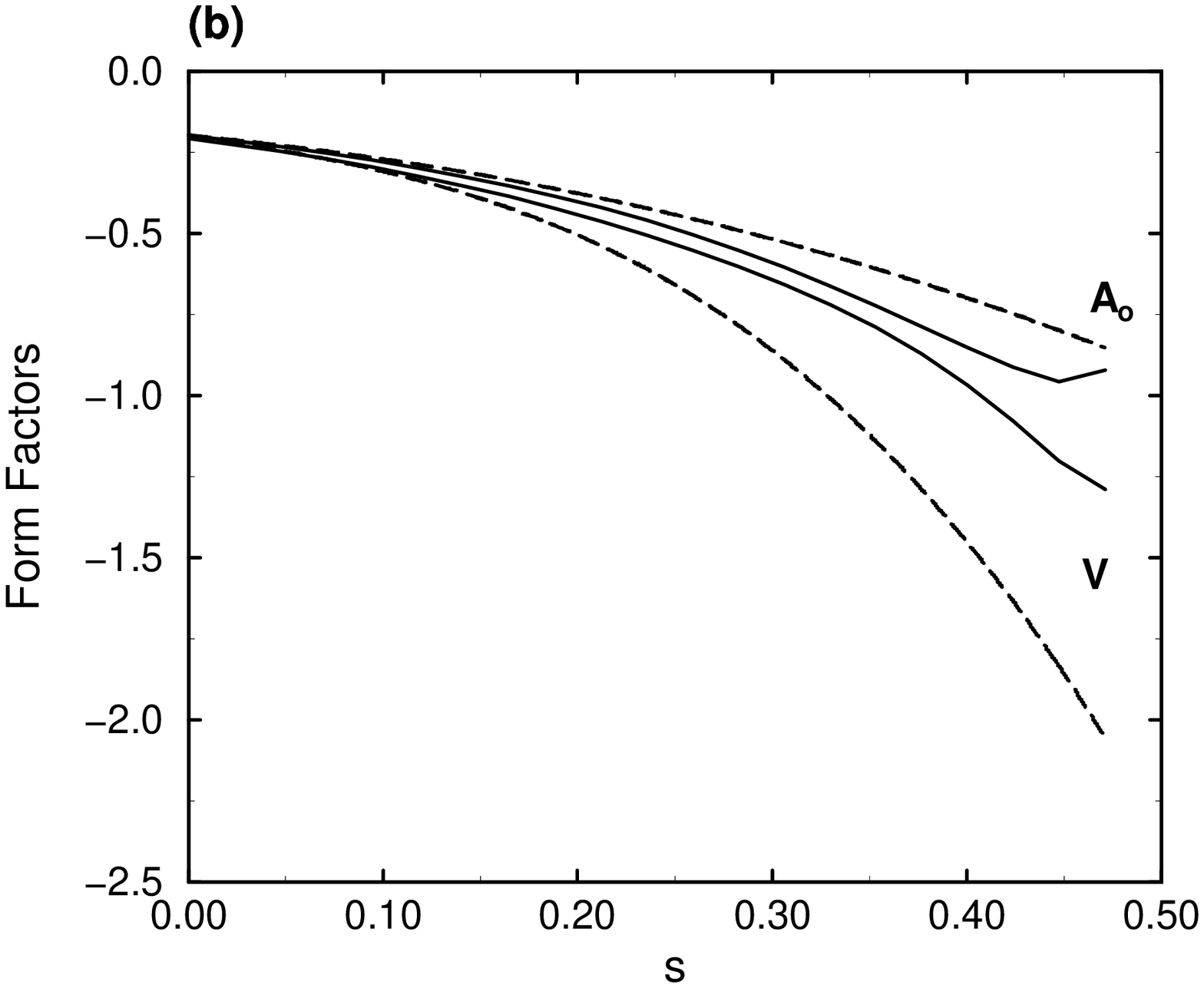} \includegraphics{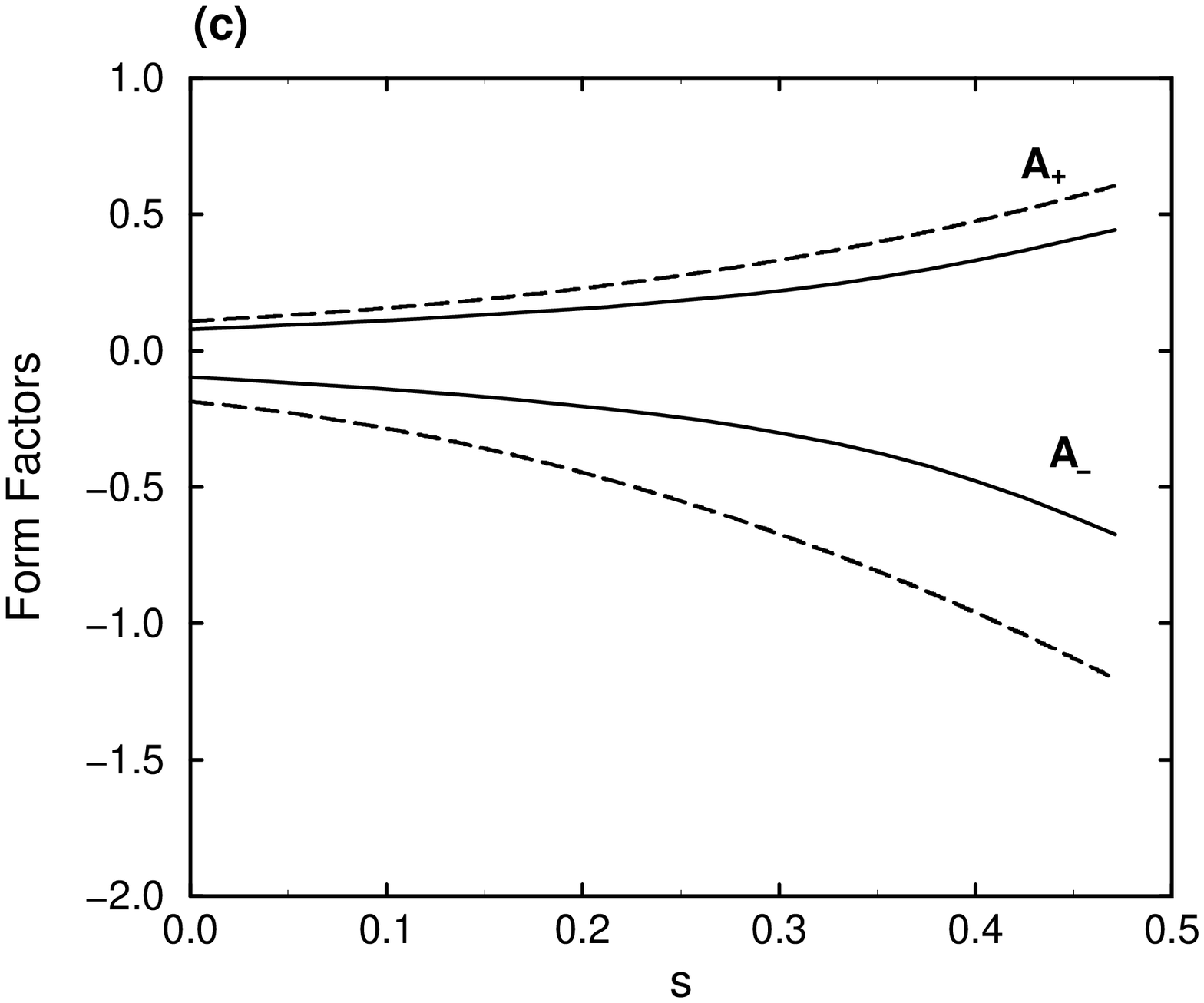}
\includegraphics{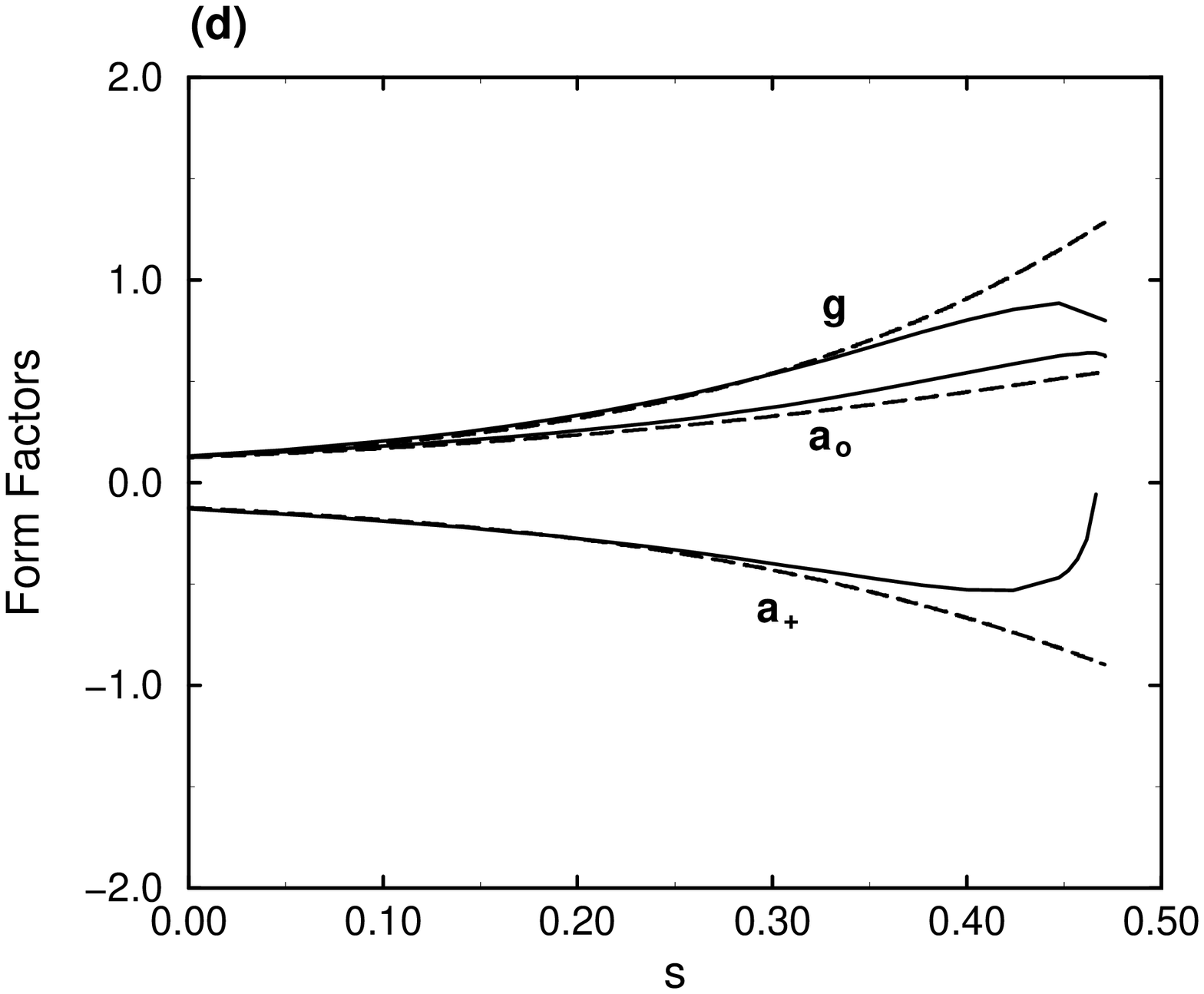} \vskip 16.5cm \caption{ Form factors of (a)
$F_{\pm,T}$ for $B_c^+\to D^+$, and (b) $V$ and $A_0$, (c) $A_{\pm}$,
and (d) $g$ and $a_{0,+}$ for $B_c^+\to D^{*+}$. The solid and dashed
curves stand for the results from the LFQM and CQM,
respectively.}
\end{figure}

\newpage
\begin{figure}[h]
\includegraphics{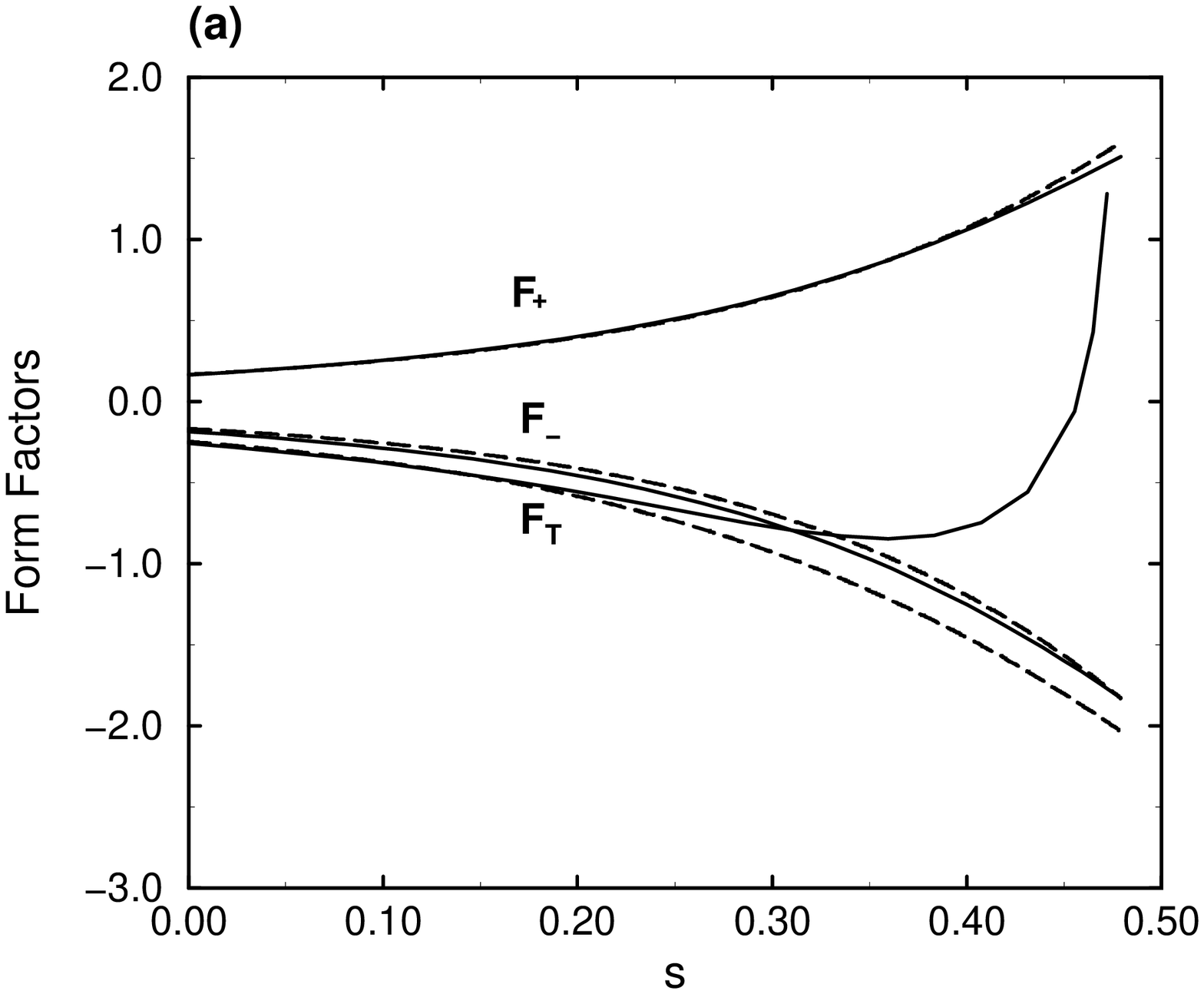} \includegraphics{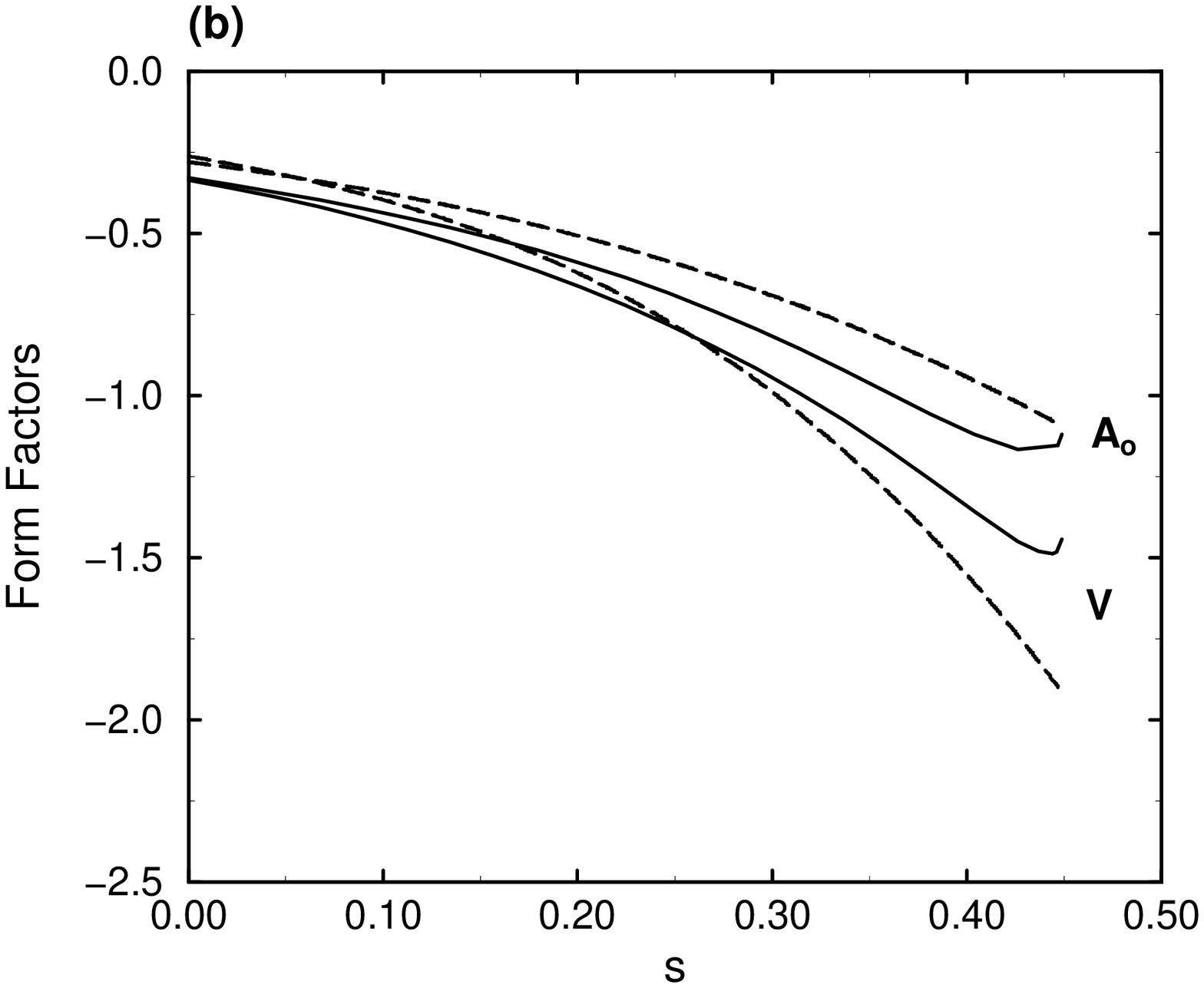} \includegraphics{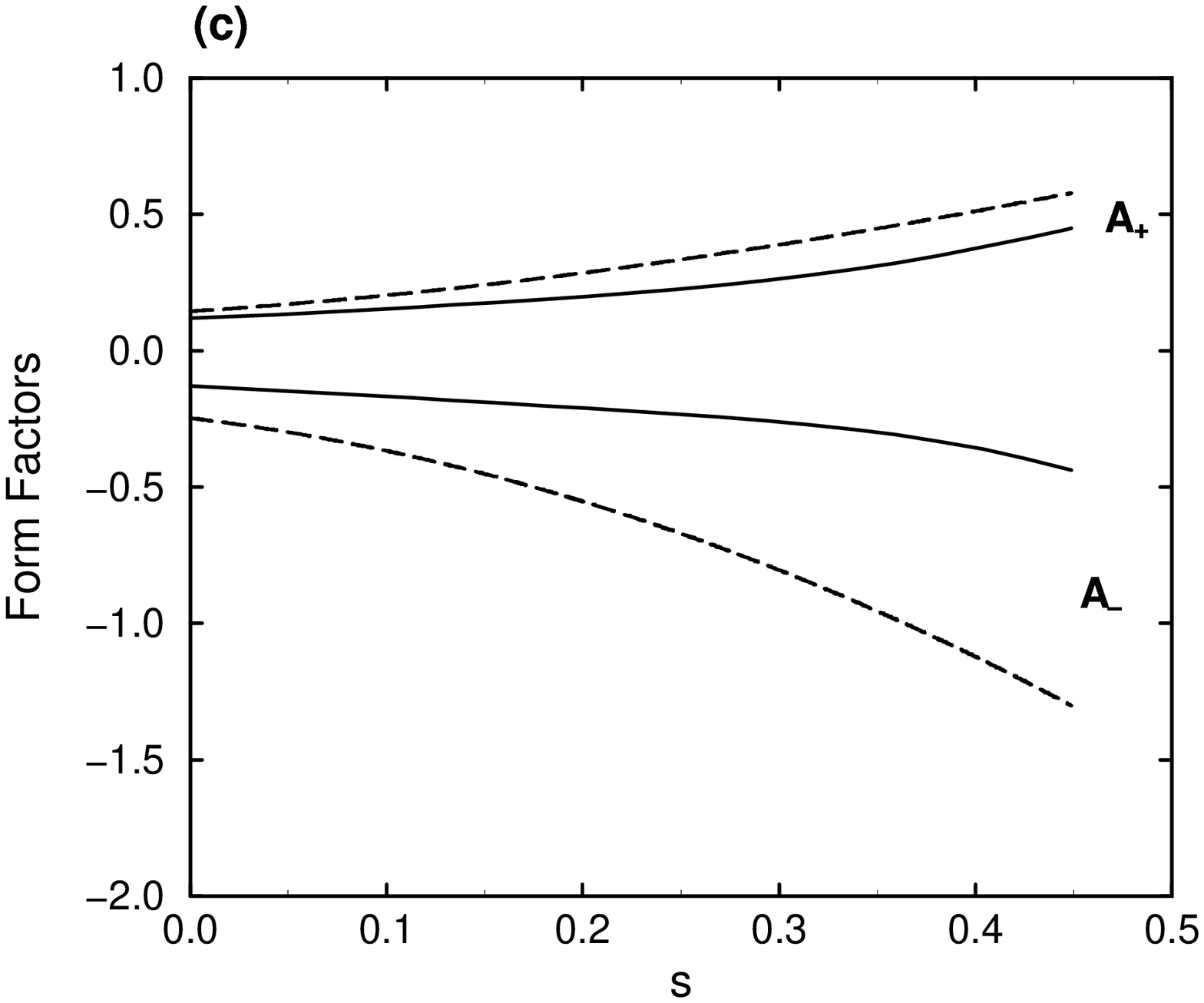} \includegraphics{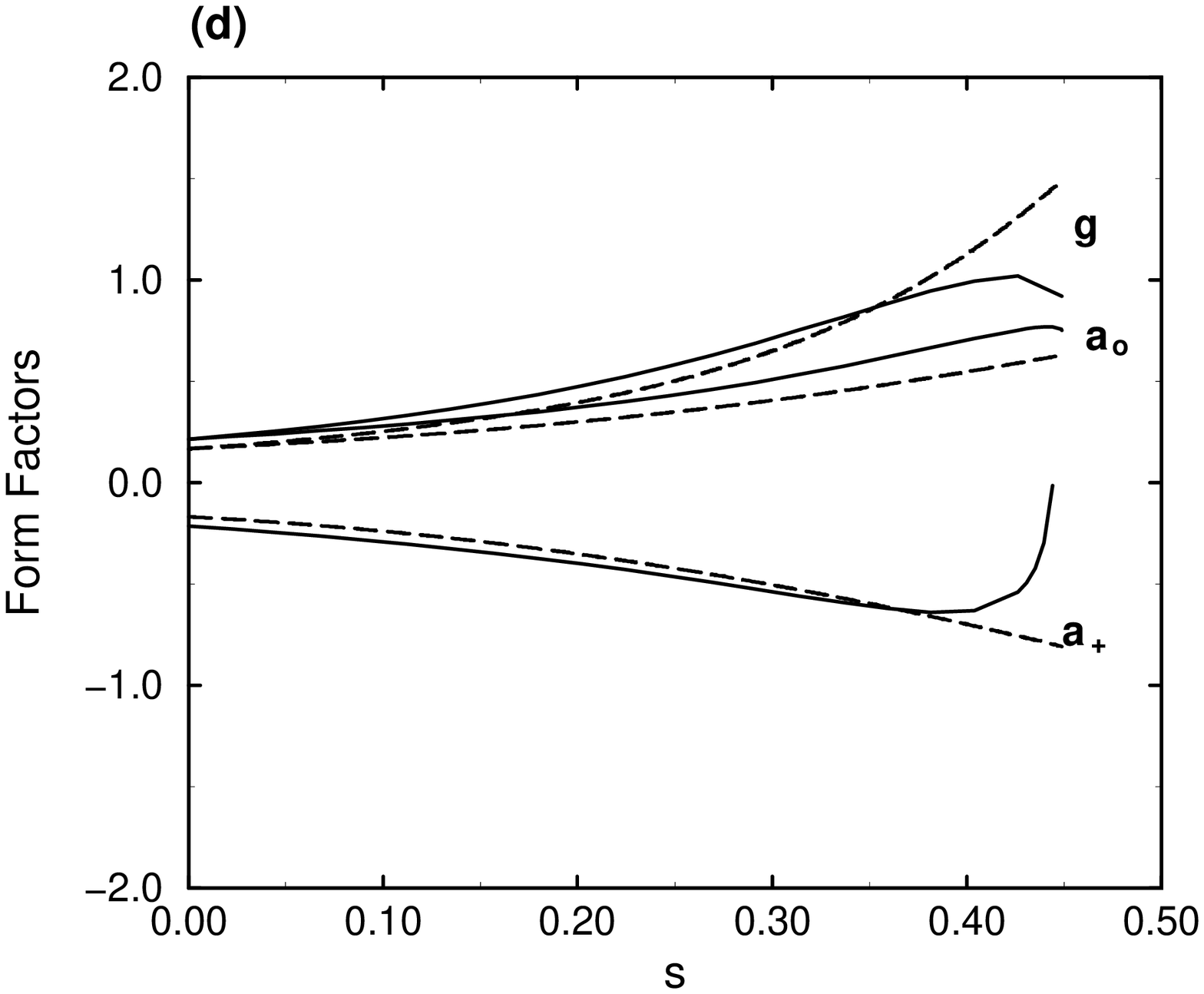} \vskip 17.cm
\caption{Same as Figure 2 but replacing $D^{(*)}$ by $D_s^{(*)}$.}
\end{figure}

\newpage
\begin{figure}[h]
\includegraphics{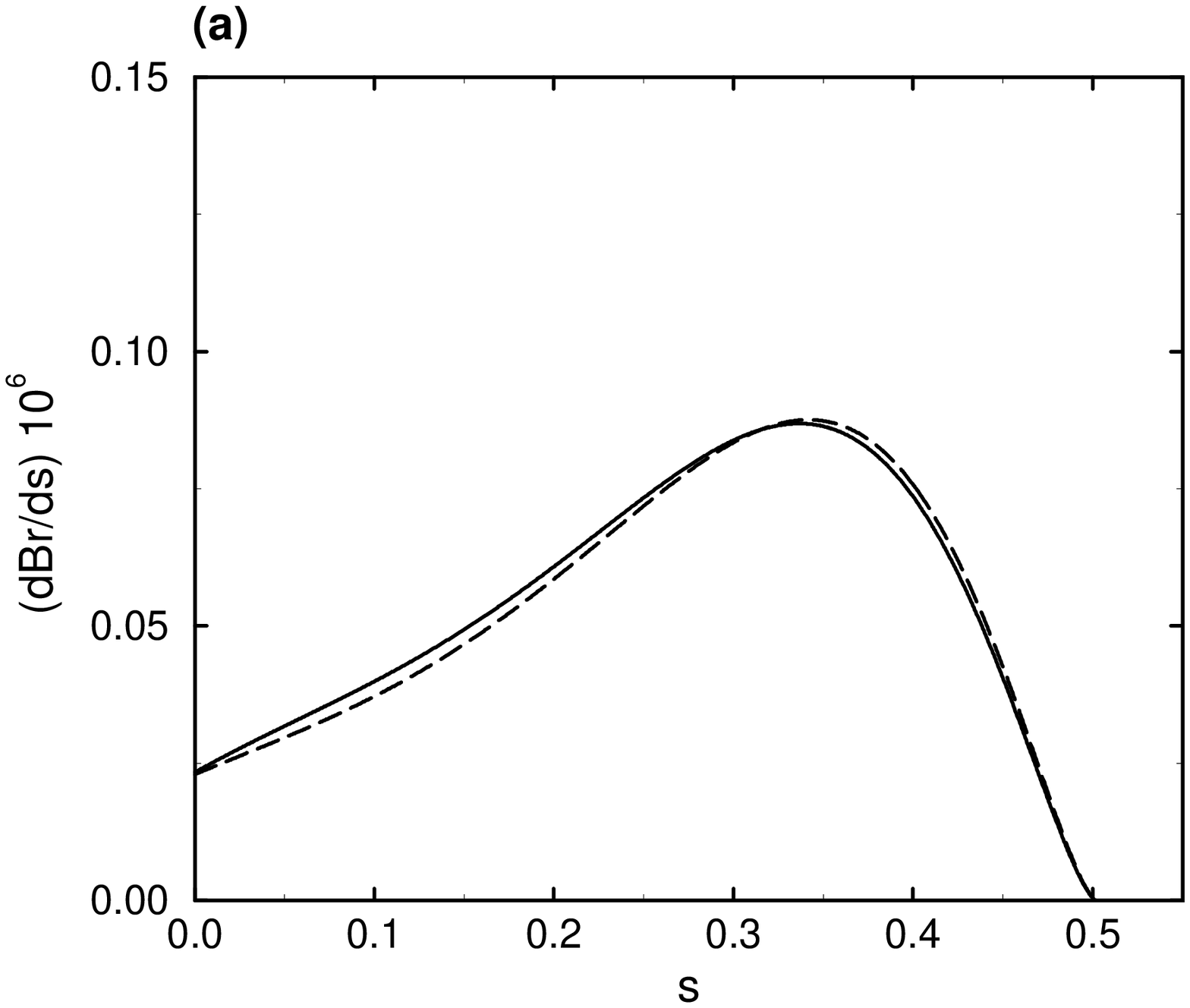} \vskip 5.cm
\end{figure}

\vskip 2.cm
\begin{figure}[h]
\includegraphics{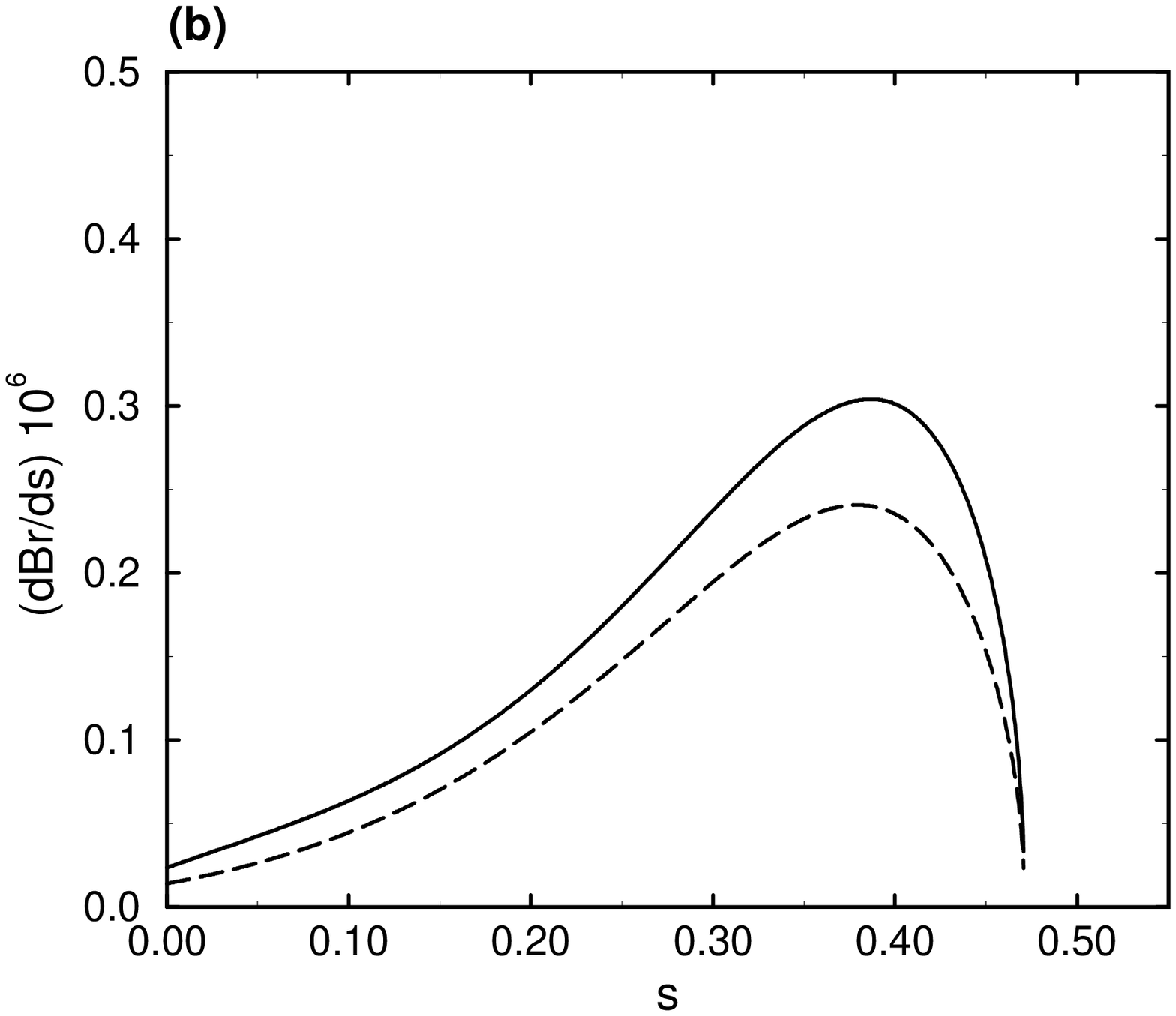} \vskip 10.5cm \caption{ Differential decay
branching ratios as a function of $s=q^{2}/m_{B_c}^{2}$ for (a)
$B_c^+\rightarrow D^+ \nu\bar{\nu} $ and (b) $B_c^+\rightarrow D^{*}
\nu\bar{\nu} $. Legend is the same as Figure 2. }
\end{figure}

\newpage
\begin{figure}[h]
\includegraphics{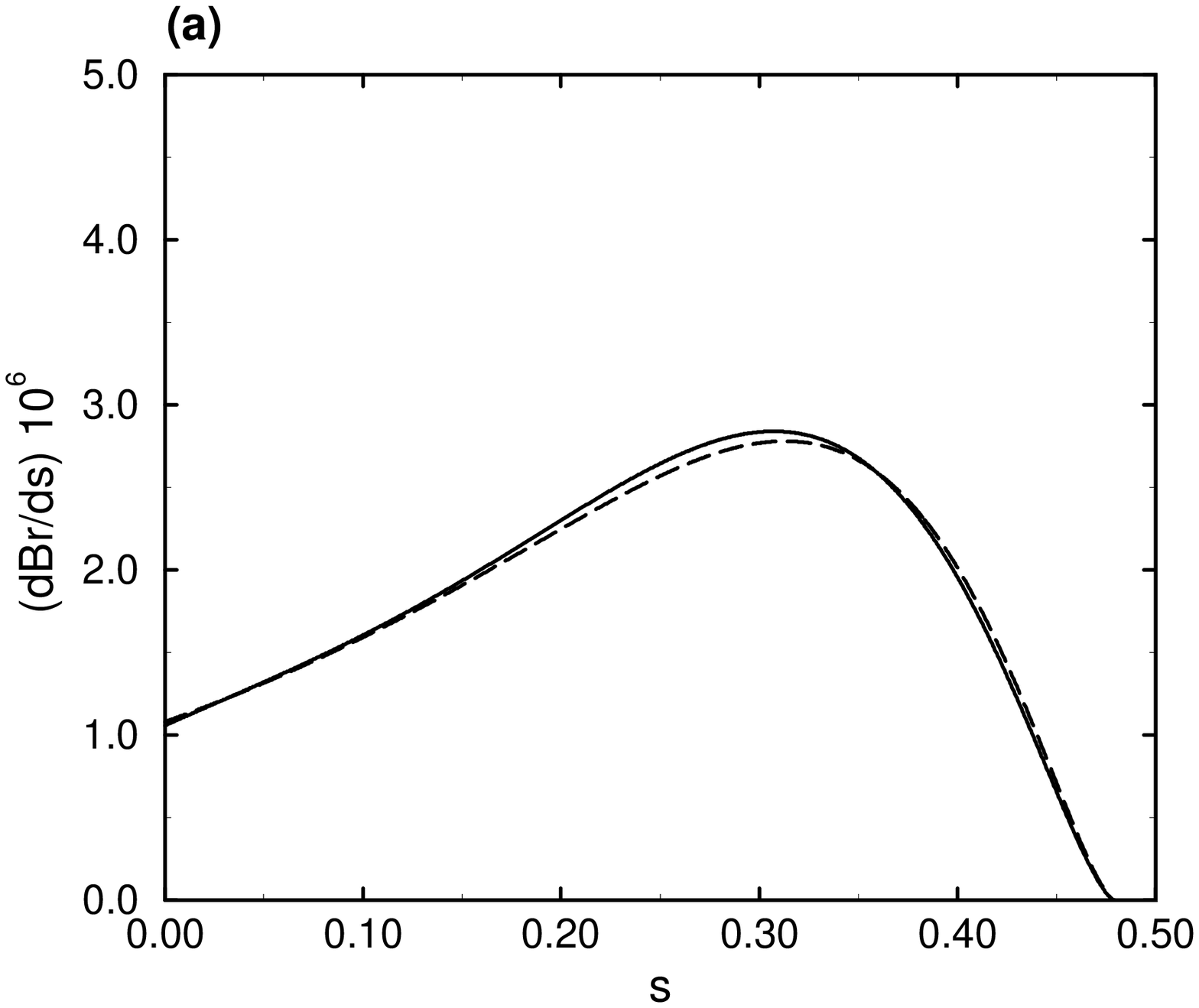} \vskip 5.cm
\end{figure}

\vskip 2.cm
\begin{figure}[h]
\includegraphics{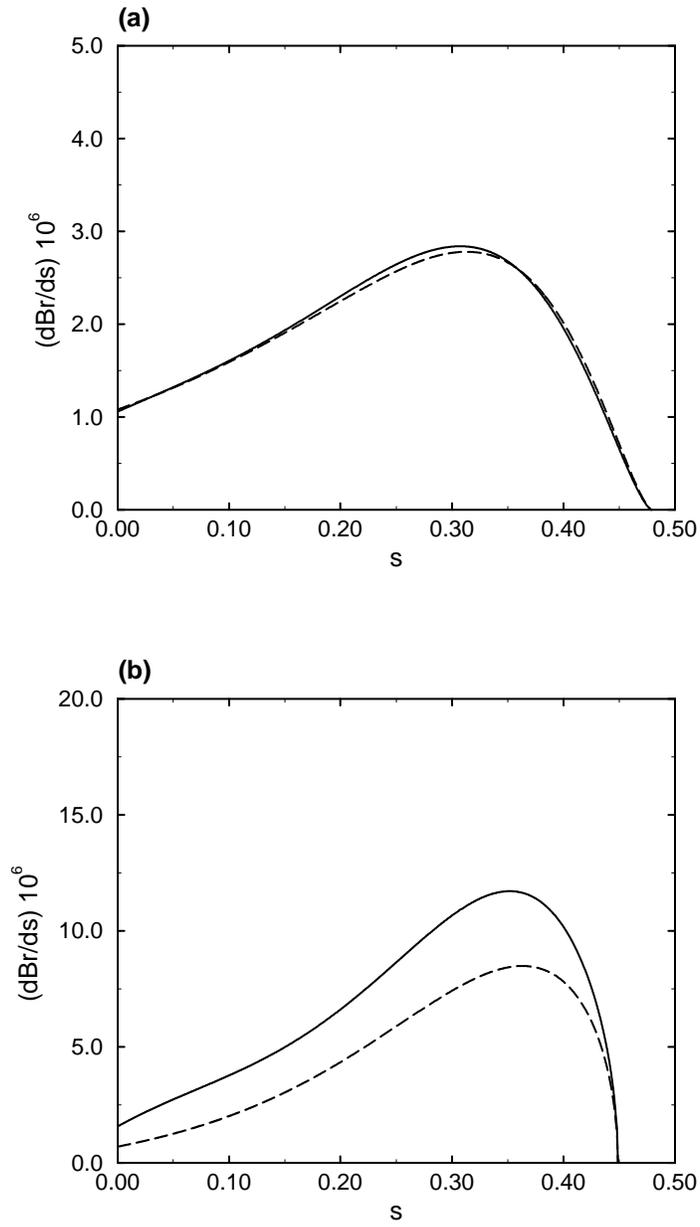} \vskip 10.5cm \caption{ Same as Figure 4 but
for (a) $B_c^+\rightarrow D_s^+\nu\bar{\nu}$ and (b) $B_c^+\rightarrow
D_s^{+*} \nu\bar{\nu}$.}
\end{figure}

\newpage
\begin{figure}[h]
\includegraphics{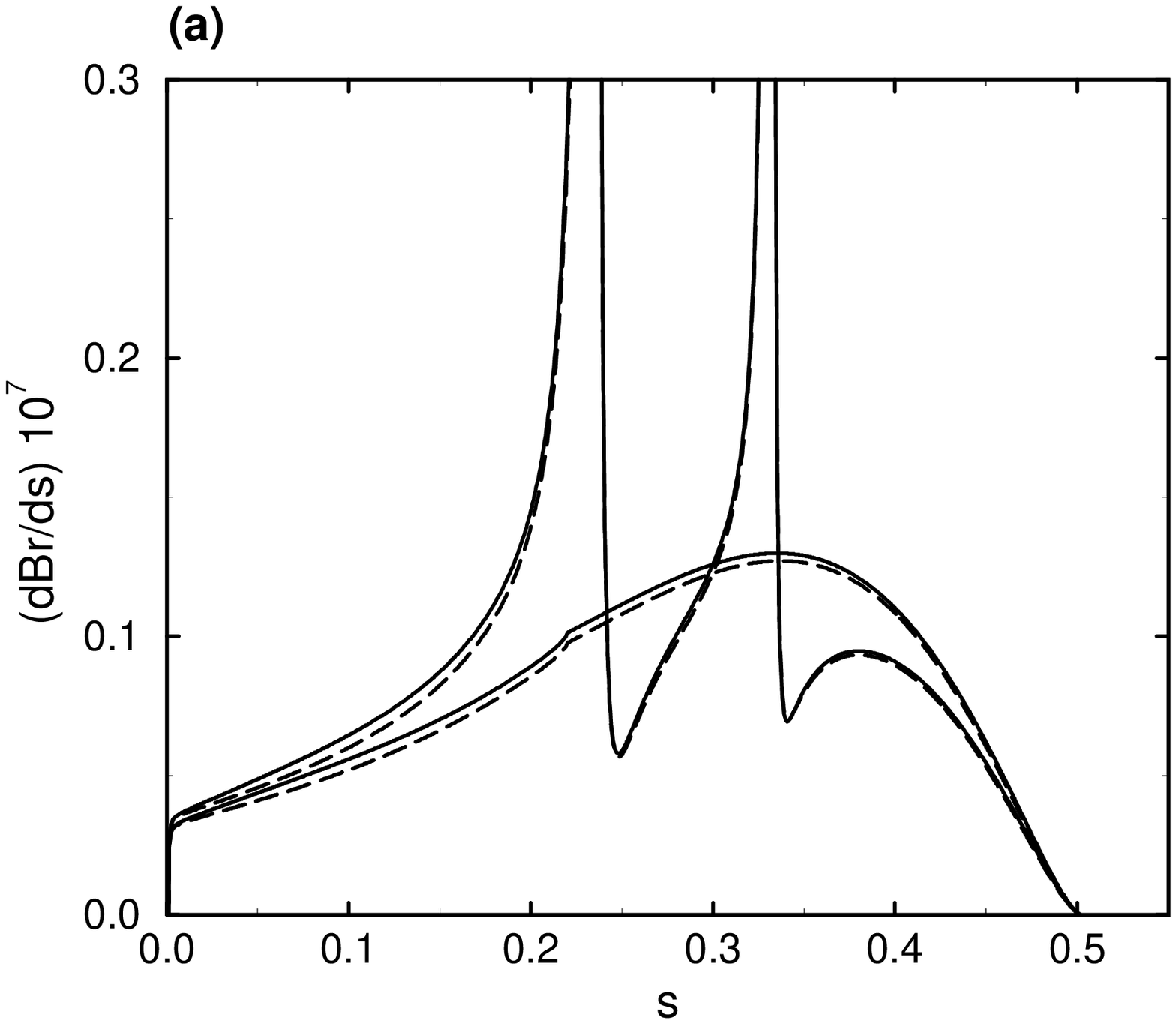} \vskip 5.cm
\end{figure}

\vskip 2.cm
\begin{figure}[h]
\includegraphics{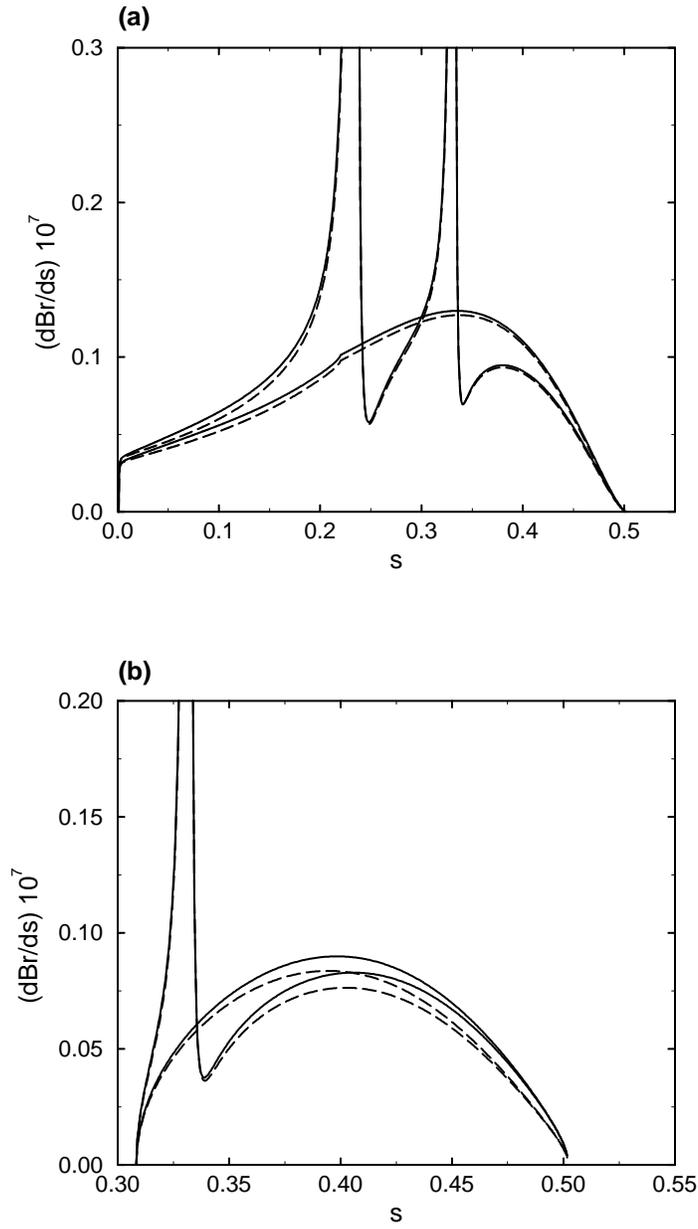} \vskip 10.5cm \caption{ Same as Figure 4 but
for (a) $B_c^+\rightarrow D^+\mu ^{+}\mu ^{-}$ and (b)
$B_c^+\rightarrow D^+ \tau ^{+}\tau ^{-}$. The curves with and
without resonant shapes represent including and non-including LD
contributions, respectively. }
\end{figure}

\newpage
\begin{figure}[h]
\includegraphics{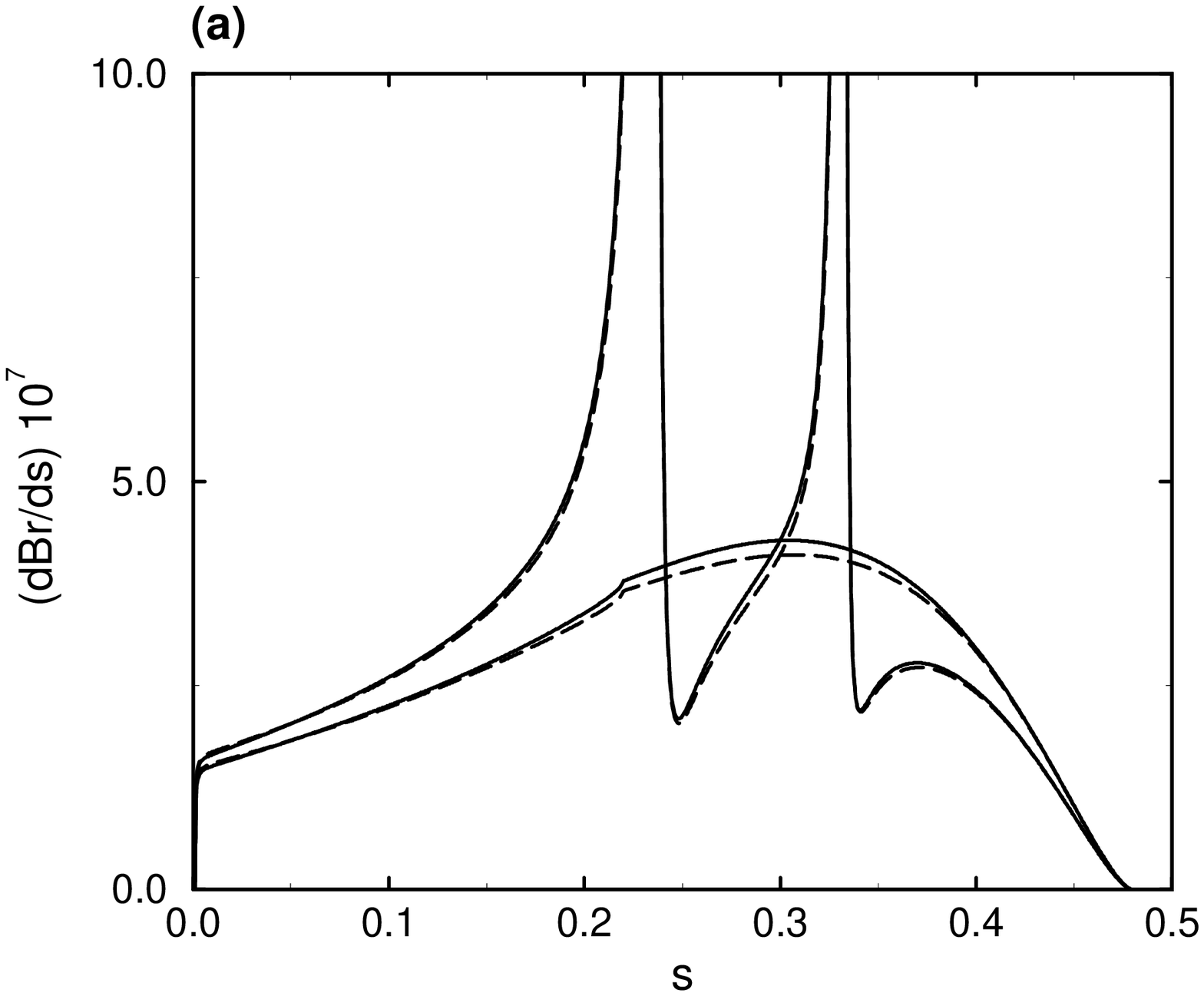} \vskip 5.cm
\end{figure}

\vskip 2.cm
\begin{figure}[h]
\includegraphics{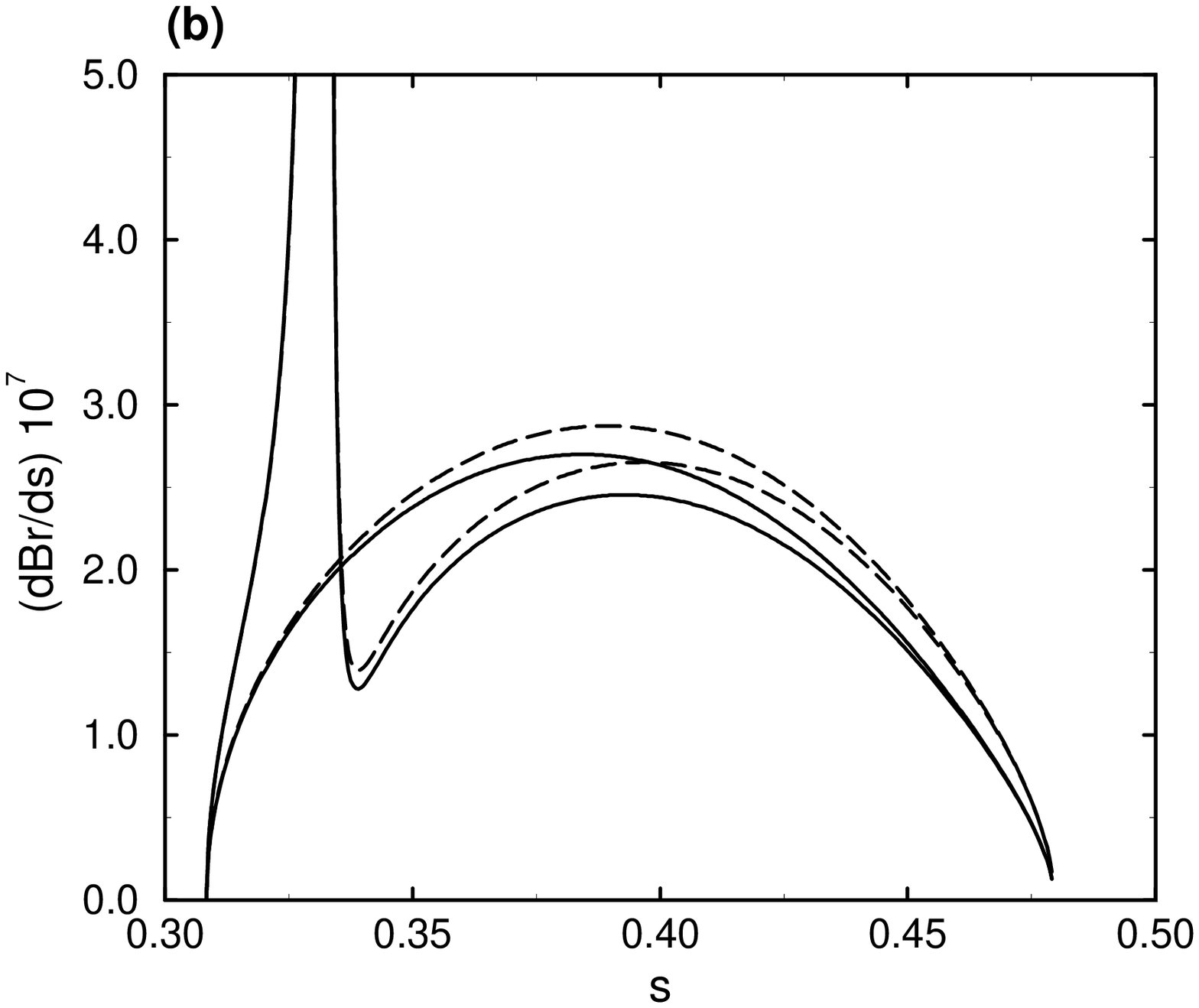} \vskip 10.5cm \caption{ Same as Figure 6 but
for (a) $B_c^+\rightarrow D_s^+ \mu ^{+}\mu ^{-}$ and (b)
$B_c^+\rightarrow D_s^+\tau^{+}\tau ^{-}$. }
\end{figure}

\newpage
\begin{figure}[h]
\includegraphics{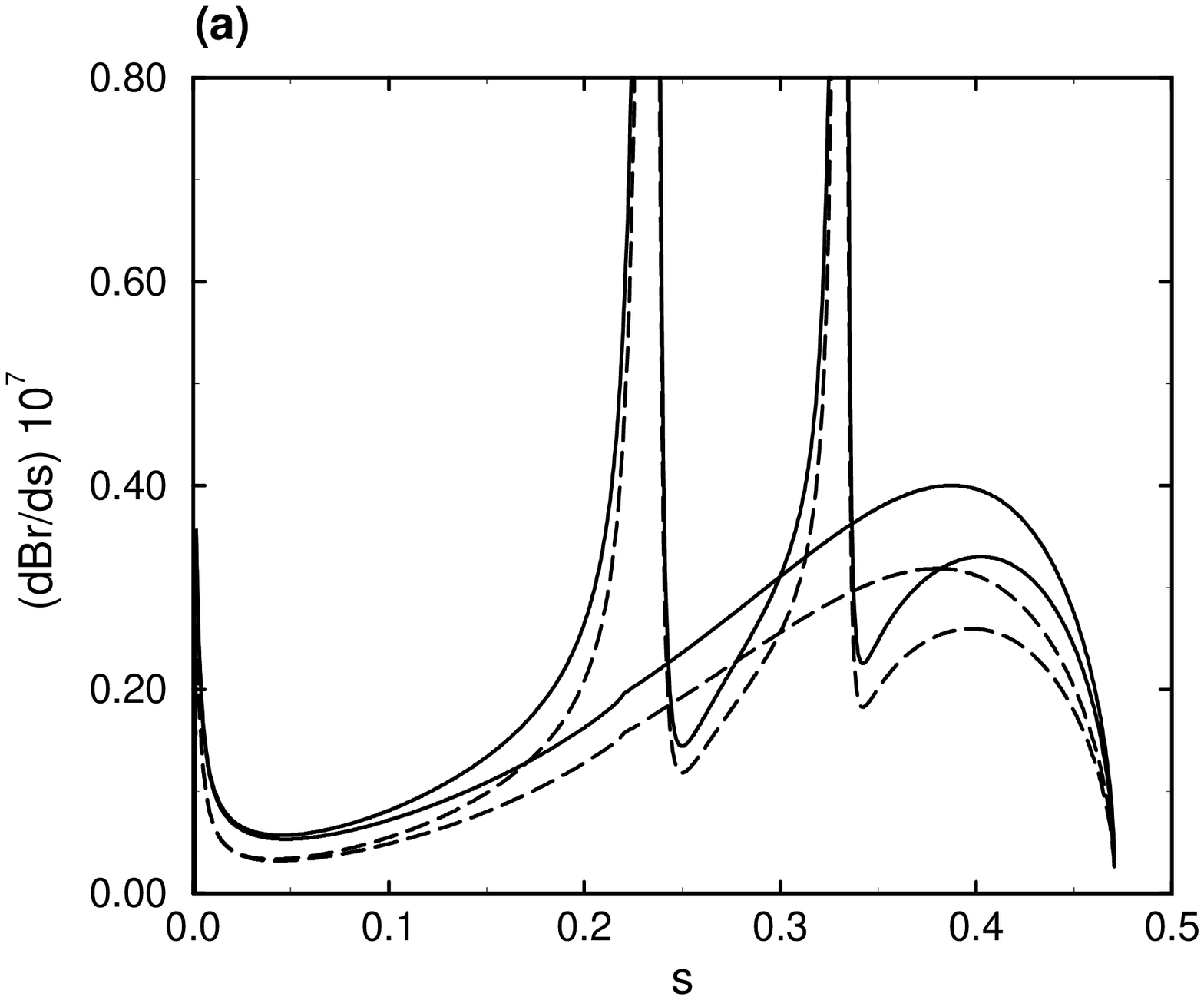} \vskip 5.cm
\end{figure}

\vskip 2.cm
\begin{figure}[h]
\includegraphics{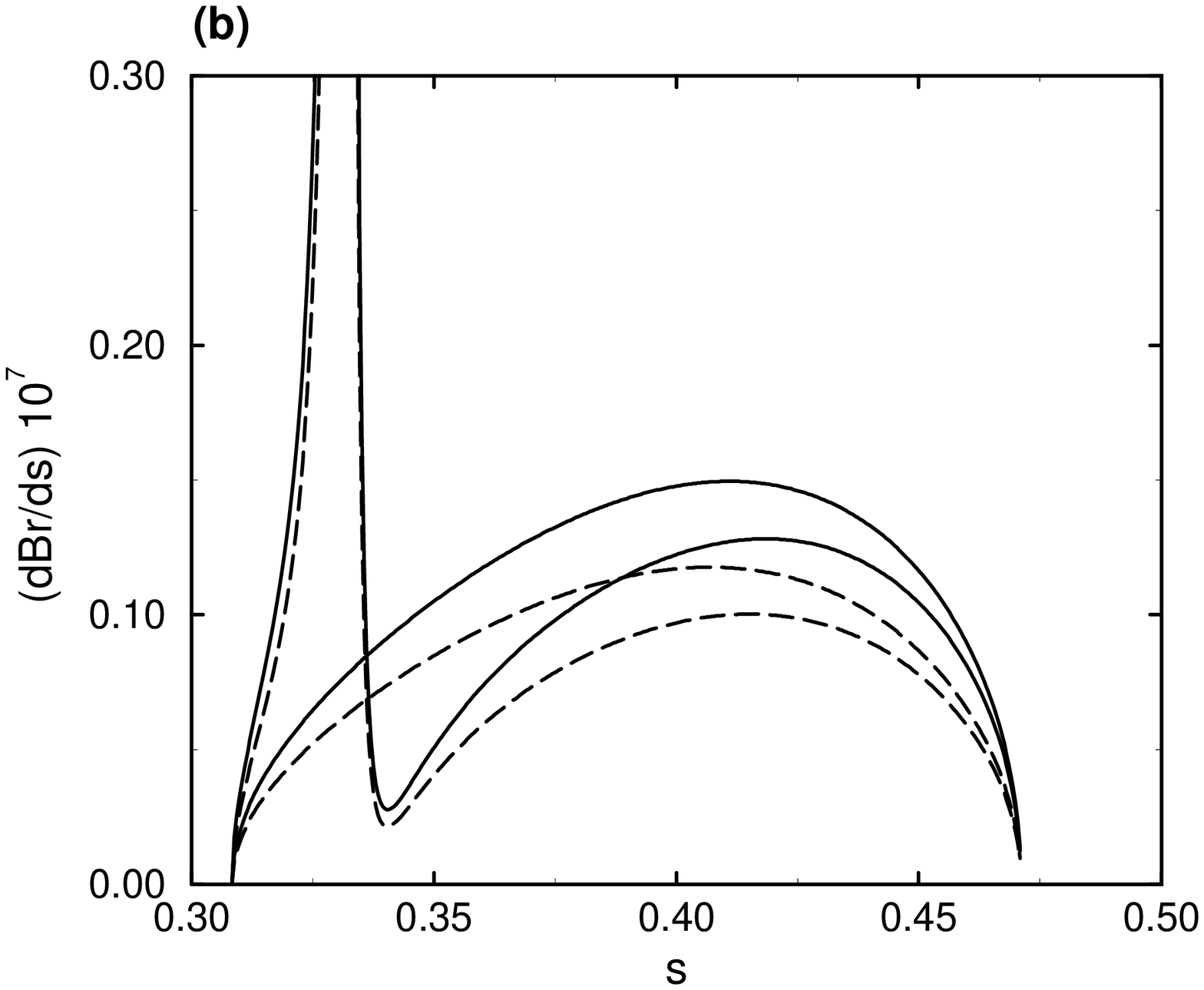} \vskip 10.5cm \caption{ Same as Figure 6 but
for (a) $B_c^+\rightarrow D^{*+} \mu ^{+}\mu ^{-}$ and (b)
$B_c^+\rightarrow D^{*+} \tau ^{+}\tau ^{-}$. }
\end{figure}

\newpage
\begin{figure}[h]
\includegraphics{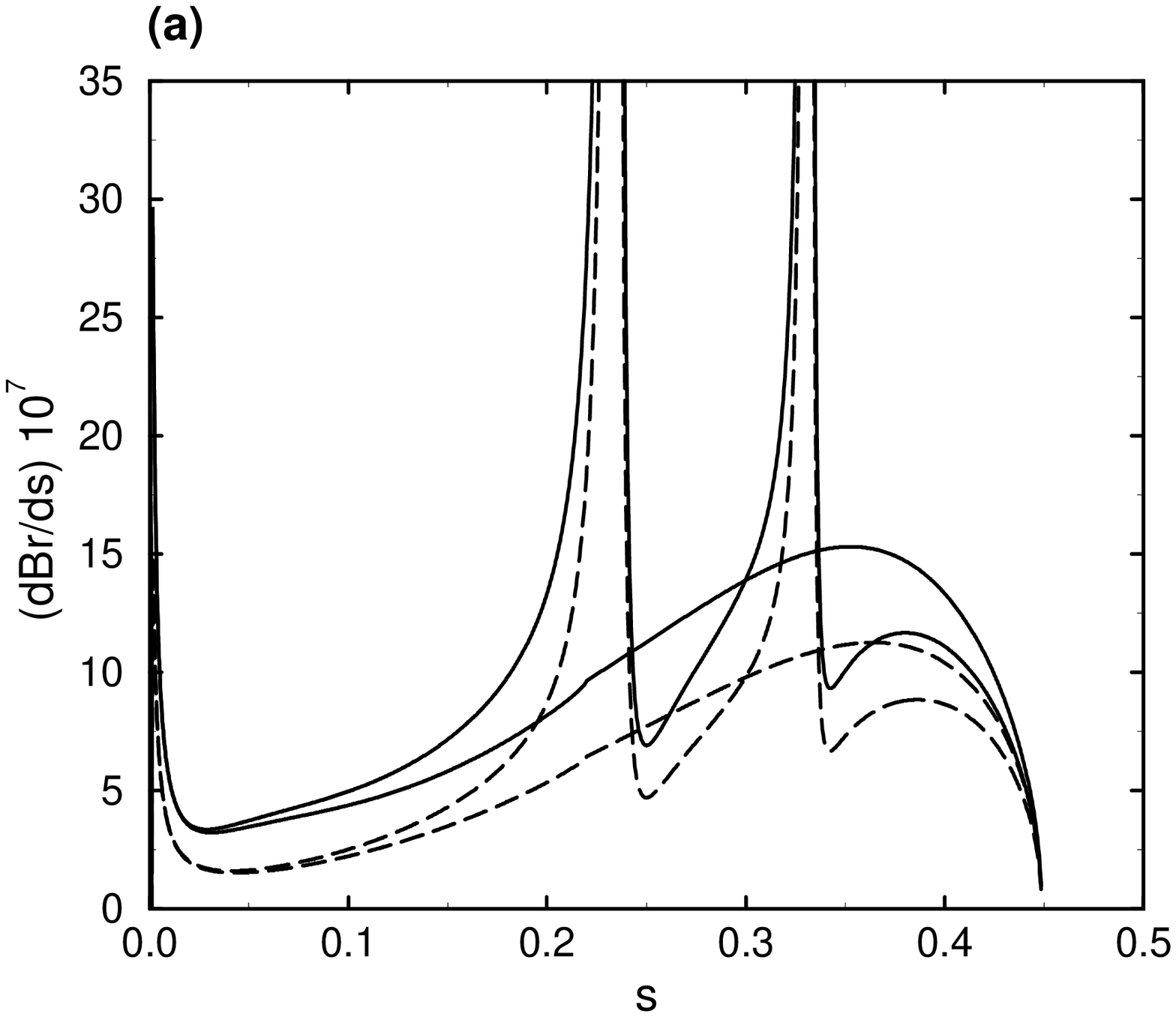} \vskip 5.cm
\end{figure}

\vskip 2.cm
\begin{figure}[h]
\includegraphics{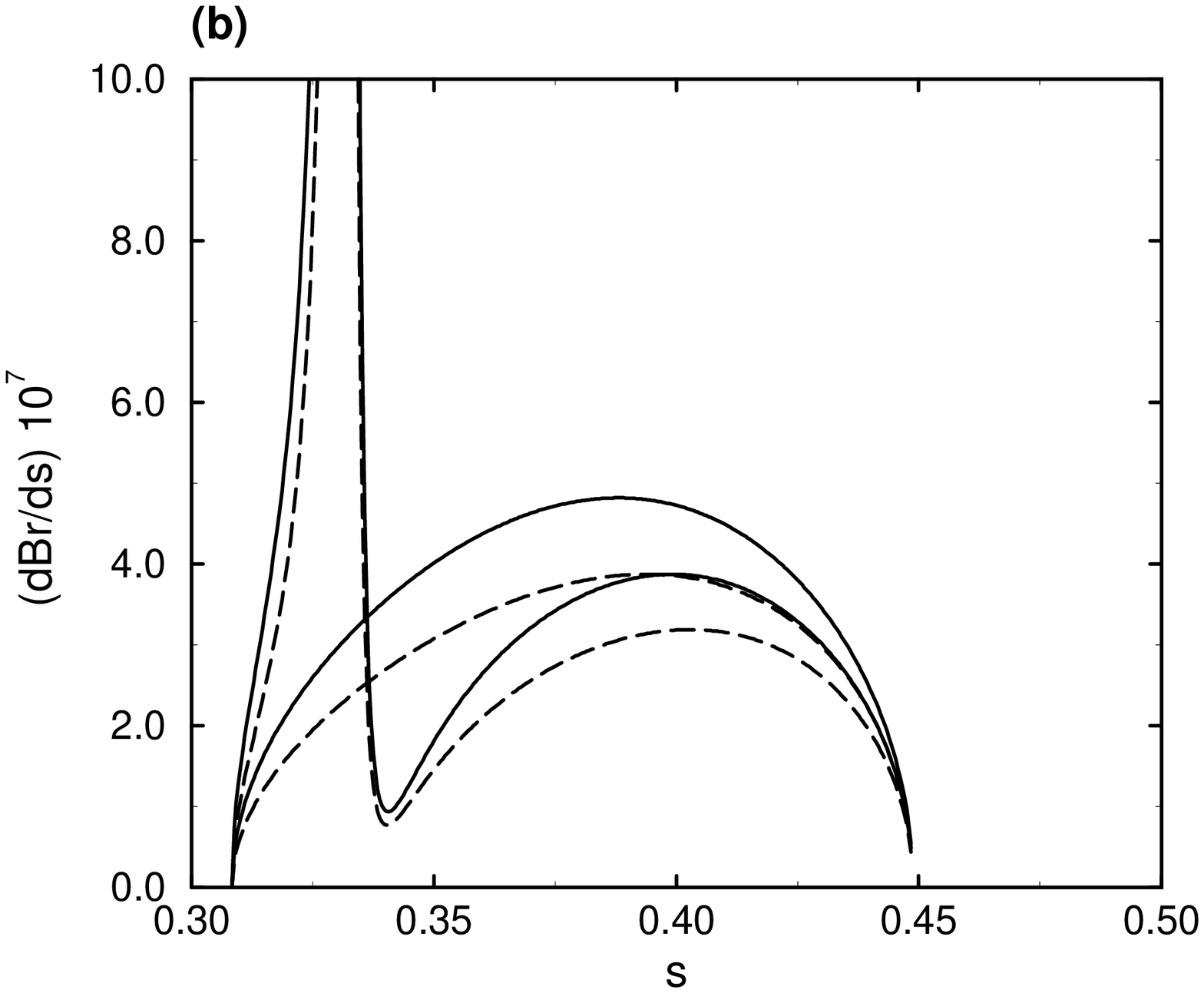} \vskip 10.5cm \caption{ Same as Figure 6 but
for (a) $B_c^+\rightarrow D_s^{*+} \mu ^{+}\mu ^{-}$ and (b)
$B_c^+\rightarrow D_s^{*+} \tau ^{+}\tau ^{-}$. }
\end{figure}

\end{document}